\documentclass[preprint,10pt]{sigplanconf}

\usepackage{balance}

\usepackage{graphicx}
\usepackage{url}

\usepackage{float}
\usepackage{wrapfig}

\usepackage{time}
\usepackage{datetime}
\usepackage{amsmath}
\usepackage{amssymb}
\usepackage{lscape}
\usepackage{time}
\usepackage{datetime}

\usepackage{color}
\makeatletter 
\let\c@lofdepth\relax 
\let\c@lotdepth\relax 
\makeatother 
\usepackage[raggedright,small,sf,SF,hang]{subfigure}

\usepackage{float}
\usepackage{wrapfig}

\usepackage{extarrows}  

\usepackage{stmaryrd}   

\begin{document}



\renewcommand{\baselinestretch}{0.92} 
\newcommand{\clpr}{CLP(${\cal R}$)}
\newcommand{\np}{${\cal NP}$~}
\newcommand{\TR}{${\cal TR}$}
\newcommand{\CTR}{${\cal CTR}$}
\newcommand{\SR}{${\cal SR}$~}
\newcommand{\del}{{\em delta}}
\newcommand{\nil}{~{\rm nil}~}
\newcommand{\I}{{\cal I}~}

\newcommand{\ignore}[1]{}

\newcommand{\stuff}[1]{
        \begin{minipage}{6in}
        {\tt \samepage
        \begin{tabbing}
        \hspace{3mm} \= \hspace{3mm} \= \hspace{3mm} \= \hspace{3mm} \= \hspace{3mm} \= \hspace{3mm} \=
\hspace{3mm} \= \hspace{3mm} \= \hspace{3mm} \= \hspace{3mm} \= \hspace{3mm} \= \hspace{3mm} \= \hspace{3mm} \= \kill
        #1
        \end{tabbing}
       }
        \end{minipage}
}

\newcommand{\mystuff}[1]{
        \begin{minipage}[b]{6in}
        {\tt \samepage
        \begin{tabbing}
        \hspace{3mm} \= \hspace{3mm} \= \hspace{3mm} \= \hspace{3mm} \= \hspace{3mm} \= \hspace{3mm} \=
\hspace{3mm} \= \hspace{3mm} \= \hspace{3mm} \= \hspace{3mm} \= \hspace{3mm} \= \hspace{3mm} \= \hspace{3mm} \= \kill
        #1
        \end{tabbing}
       }
        \end{minipage}
}

\newcommand{\newmystuff}[1]{
        \begin{minipage}[b]{6in}
        {\tt \samepage
        \begin{tabbing}
        \hspace{3mm} \= \hspace{3mm} \= \hspace{3mm} \= \hspace{3mm} \= \hspace{3mm} \= \hspace{3mm} \=
\hspace{3mm} \= \hspace{3mm} \= \hspace{3mm} \= \hspace{3mm} \= \hspace{3mm} \= \hspace{3mm} \= \hspace{3mm} \= \kill
        #1
        \end{tabbing}
       }
        \end{minipage}
}

\newcommand{\Rule}[2]{\genfrac{}{}{0.5pt}{}
{{\setlength{\fboxrule}{0pt}\setlength{\fboxsep}{3mm}\fbox{$#1$}}}
{{\setlength{\fboxrule}{0pt}\setlength{\fboxsep}{3mm}\fbox{$#2$}}}}

\newcommand{\rct}{\mbox{\it clp}}
\newcommand{\lif}{\ {\tt:\!\!-} \ }
\newcommand{\ctbl}{\,{\mbox{\footnotesize$|$}\tt-\!\!\!\!>}\,}
\newlength{\colwidth}
\setlength{\colwidth}{.47\textwidth}


\newcommand{\eqdef}{\stackrel{\rm def}{=}}
\newcommand{\sys}{\mbox{\bf sys}}
\newcommand{\var}{\mbox{\frenchspacing \it var}}
\newcommand{\lfp}{\mbox{\it \frenchspacing lfp}}
\newcommand{\gfp}{\mbox{\it \frenchspacing gfp}}
\newcommand{\assert}{\mbox{\it \frenchspacing assert}}
\newcommand{\buffer}{\mbox{\it \frenchspacing Buffer}}
\newcommand{\clpif}{\mbox{\tt :-}}


\newtheorem{definition}{Definition}
\newtheorem{theorem}{Theorem}
\newtheorem{lemma}{Lemma}
\newtheorem{proposition}{Proposition}
\newtheorem{corollary}{Corollary}
\newtheorem{example}{Example}
\newtheorem{myproof}{Proof Outline}

\newtheorem{assumption}{Assumption}

\newcommand{\QED}{\nolinebreak\hskip 1em
        \framebox[0.5em]{\rule{0ex}{1.0ex}}}

\newcommand{\close}{{\frenchspacing close }}
\newcommand{\rules}{{\frenchspacing rules }}
\newcommand{\lhs}{{\frenchspacing lhs }}
\newcommand{\rhs}{{\frenchspacing rhs }}
\newcommand{\wrt}{{\frenchspacing wrt. }}
\newcommand{\fold}{\mbox{\it\frenchspacing old}}
\newcommand{\exactunfold}{\mbox{\it\frenchspacing exactunfold}}

\newcommand{\mgu}{{\frenchspacing mgu }}
\newcommand{\size}{{\frenchspacing size }}
\newcommand{\obs}{{\frenchspacing obs }}
\newcommand{\trace}{{\frenchspacing trace }}

\newcommand{\xxx}{\mbox{\Large $\vartriangleright$}}

\newcommand{\undeniable}{\mbox{\Large $\vartriangleright$}\hspace{-8pt}\raisebox{2pt}{\tiny u}~~}
\newcommand{\inevitable}{\mbox{\Large $\vartriangleright$}\hspace{-8pt}\raisebox{2pt}{\tiny i}~~\,}

\newcommand{\reachable}{\mbox{\Large $\vartriangleright$}\hspace{-9pt}\raisebox{1pt}{\tiny *}~~\,}

\newcommand{\lbr}{\mbox{$[\![$}}
\newcommand{\rbr}{\mbox{$]\!]$}}
\newcommand{\ct}[1]{\mbox{\lbr$\vec{#1}$\rbr}}
\newcommand{\cs}[1]{\mbox{\lbr#1\rbr}}
\newcommand{\clp}[1]{{\it \frenchspacing clp}(\mbox{$#1$})}
\newcommand{\pp}[1]{\mbox{{\color{blue}{$\langle$#1$\rangle$}}}}
\newcommand{\sst}[3]{\mbox{${\color{magenta} \langle} #1, #2, #3 {\color{magenta} \rangle}$}}

\newcommand{\ao}{\mbox{${\cal A}$}}
\newcommand{\co}{\mbox{${\cal C}$}}
\newcommand{\po}{\mbox{${\cal P}$}}

\newcommand{\tA}{{\tilde{A}}}
\newcommand{\tB}{{\tilde{B}}}
\newcommand{\tC}{{\tilde{C}}}

\newcommand{\A}{\mbox{$\cal A$}}
\newcommand{\B}{\mbox{$\cal B$}}
\newcommand{\D}{\mbox{$\cal D$}}
\newcommand{\E}{\mbox{$\cal E$}}
\newcommand{\M}{\mbox{$\cal M$}}
\newcommand{\V}{\mbox{$\cal V$}}
\newcommand{\unfold}{\mbox{\sc unfold}}

\newcommand{\G}{\mbox{$\cal G$}}
\newcommand{\Gone}{\mbox{${\cal G}_{\!1}$}}
\newcommand{\Gi}{\mbox{${\cal G}_{\!i}$}}
\newcommand{\Gn}{\mbox{${\cal G}_{\!n}$}}

\newcommand{\GL}{\mbox{${\cal G}_{\!L}$}}
\newcommand{\GR}{\mbox{${\cal G}_{\!R}$}}

\newcommand{\HH}{\mbox{$\cal H$}}
\newcommand{\HHone}{\mbox{${\cal H}_{\!1}$}}
\newcommand{\HHi}{\mbox{${\cal H}_{\!i}$}}
\newcommand{\HHj}{\mbox{${\cal H}_{\!j}$}}
\newcommand{\HHn}{\mbox{${\cal H}_{\!n}$}}
\newcommand{\HHm}{\mbox{${\cal H}_{\!m}$}}

\newcommand{\Mone}{\mbox{${\cal M}_{\!1}$}}
\newcommand{\Mtwo}{\mbox{${\cal M}_{\!2}$}}
\newcommand{\MMi}{\mbox{${\cal M}_{\!i}$}}
\newcommand{\MMn}{\mbox{${\cal M}_{\!n}$}}
\newcommand{\MMnone}{\mbox{${\cal M}_{\!n-1}$}}

\newcommand{\PsiL}{\mbox{$\Psi_{\!L}$}}
\newcommand{\PsiR}{\mbox{$\Psi_{\!R}$}}
\newcommand{\PsiB}{\mbox{$\Psi_{\!B}$}}

\newcommand{\Af}{A_{\!f}}
\newcommand{\Ax}{A_{\!1}}
\newcommand{\Nf}{N_{\!f}}
\newcommand{\Xf}{X_{\!f}}
\newcommand{\Yf}{Y_{\!f}}
\newcommand{\Hxxx}{H_{\!f}}

\newcommand{\cts}{\mbox{$\mapsto$}}
\newcommand{\sepimp}{\mbox{$-\!*$}}

\newcommand{\arr}[3]{\mbox{$\langle \mbox{#1,#2,#3} \rangle$}}
\newcommand{\triple}[3]{\langle{#1},{#2},{#3}\rangle}
\newcommand{\aquadruple}[4]{\langle{#1},{#2},{#3},{#4}\rangle}
\newcommand{\baeq}[2]{\mbox{$=_{\[#1 .. #2\]}$}}

\newcommand{\ite}[3]{\mbox{\textit{ite}}({#1},{#2},{#3})}
\newcommand{\pred}[1]{\mbox{\textit{#1}}}
\newcommand{\ptab}{~~~~}

\newlength{\vitelen}
\settowidth{\vitelen}{\mbox{\textit{ite}}(}
\newcommand{\vite}[3]{\begin{array}[t]{l}\mbox{\textit{ite}}({#1},\\
\hspace{\vitelen}{#2},\\
\hspace{\vitelen}{#3}
    \end{array}}

\newcommand{\Hfx}{H_{\!f}}
\newcommand{\Hx}{H_{\!1}}
\newcommand{\Hxx}{H_{\!2}}
\newcommand{\Pf}{P_{\!f}}
\newcommand{\Pz}{P_{\!0}}
\newcommand{\Px}{P_{\!1}}

\newcommand{\Ix}{I_{\!1}}
\newcommand{\Ixx}{I_{\!2}}
\newcommand{\Jf}{J_{\!f}}
\newcommand{\Jx}{J_{\!1}}
\newcommand{\Jxx}{J_{\!2}}

\newcommand{\witness}[1]{\mbox{$\omega_{#1}$}}
\newcommand{\awitness}[1]{\mbox{$\sigma_{#1}$}}
\newcommand{\transformer}{\mbox{$\Delta$}}

\newcommand{\func}[1]{\mbox{\textsf{#1}}}

\floatstyle{boxed}
\restylefloat{figure}

%

\renewcommand\labelitemi{$\bullet$}
\renewcommand\labelitemii{\normalfont\bfseries --}

\renewcommand\floatpagefraction{.9}
\renewcommand\topfraction{.9}
\renewcommand\bottomfraction{.9}
\renewcommand\textfraction{.1}   
\setcounter{totalnumber}{50}
\setcounter{topnumber}{50}
\setcounter{bottomnumber}{50}

\raggedbottom

\makeatletter
\newcommand{\manuallabel}[2]{\def\@currentlabel{#2}\label{#1}}
\makeatother

\newcounter{chapcount}
\newcommand{\chapcountreset}{\setcounter{chapcount}{0}}
\chapcountreset

\newcounter{excount}
\newcommand{\exreset}{\setcounter{excount}{0}}
\exreset
\newcommand{\newexample}[1]{\addtocounter{excount}{1}
{\vspace{2mm} \noindent \mbox{{\scriptsize EXAMPLE} \examplecount~\emph{#1}:}}}

\newcommand{\examplecount}{\arabic{excount}}

\newcommand{\exlabel}[1]{\manuallabel{#1}{\examplecount}}

\newcommand{\todo}[1]{
\noindent \textbf{{\huge TODO:} #1}
}

\newcommand{\algocombine}{\textsc{Combine}}
\newcommand{\algoincr}{\textsc{IncrementalAnalysis}}
\newcommand{\algoai}{\textsc{AbstractInterpretation}}
\newcommand{\algorefine}{\textsc{RefineUnfold}}
\newcommand{\algopropagate}{\textsc{PropagateBack}}
\newcommand{\algonondom}{\textsc{NonDominatedAI}}
\newcommand{\algorefineheu}{\emph{RefinementHeuristic}}
\newcommand{\algoboundsheu}{\emph{BoundsHeuristic}}
\newcommand{\algocoincide}{\textsc{BoundsGoodEnough}}
\newcommand{\algowitness}{\textsc{Witnesses}}
\newcommand{\algowitnessheu}{\emph{WitnessHeuristic}}
\newcommand{\UB}{\mathcal{U}}
\newcommand{\LB}{\mathcal{L}}

\newcommand{\INFEASIBLE}{{\sf INFEASIBLE}}
\newcommand{\SUBSUMES}{{\sf SUBSUMES}}
\newcommand{\TERMINAL}{{\sf TERMINAL}}
\newcommand{\SYMSTEP}{{\sf SYMSTEP}}
\newcommand{\DOMINATES}{{\sf DOMINATES}}
\newcommand{\INTP}{{\sf INTP}}

\newcommand{\algorithmicinput}{\textbf{Input:~}}
\newcommand{\algorithmicoutput}{\textbf{Output:~}}
\newcommand{\algorithmicglobal}{\textbf{Globally:~}}
\newcommand{\algorithmicfunction}{\textbf{Function}\ }
\newcommand{\algorithmicfunctionend}{\textbf{EndFunction}\ }
\newcommand{\memoed}{\mbox{\func{memoed}}}
\newcommand{\memoize}{\mbox{\func{memoize}}}

\newcommand{\memotable}{\mbox{\pred{Table}}}
\newcommand{\wpc}{\mbox{\textsf{$\widehat{wlp}$}}}

\newcommand{\tracer}{\mbox{\sc tracer}}

\newcommand{\interp}{\mbox{$\pred{Intp}$}}
\newcommand{\emanate}{\mbox{\func{outgoing}}}
\newcommand{\absiteration}{\mbox{\func{loop\_end}}}
\newcommand{\step}{\mbox{\func{TransStep}}}
\newcommand{\compress}{\mbox{\func{JoinVertical}}}
\newcommand{\join}{\mbox{\func{JoinHorizontal}}}

\newcommand{\program}{\mbox{$\cal P$}}
\newcommand{\N}{\mbox{$\cal N$}}

\newcommand{\Assign}{\mbox{:=}}
\newcommand{\pair}[2]{\langle{#1},{#2}\rangle}
\newcommand{\tuple}[3]{\langle{#1},{#2},{#3}\rangle}
\newcommand{\quadruple}[4]{\langle{#1},{#2},{#3},{#4}\rangle}
\newcommand{\quintuple}[5]{[{#1},{#2},{#3},{#4},{#5}]}
\newcommand{\states}{\mbox{$\mathcal{S}$}}

\newcommand{\sat}{{\small \textsf{SAT}}}
\newcommand{\unsat}{\textsf{UNSAT}}
\newcommand{\smt}{{\small \textsf{SMT}}}
\newcommand{\por}{{\small \textsf{POR}}}
\newcommand{\dpor}{{\small \textsf{DPOR}}} 
\newcommand{\si}{{\small \textsf{SI}}}
\newcommand{\ti}{{\small \textsf{TI}}}
\newcommand{\cegar}{{\small \textsf{CEGAR}}}
\newcommand{\al}{{\small \textsf{AL}}}
\newcommand{\rcsp}{{\small \textsf{RCSP}}}
\newcommand{\wcet}{{\small \textsf{WCET}}}
\newcommand{\ilp}{{\small \textsf{ILP}}}
\newcommand{\ipet}{{\small \textsf{IPET}}}
\newcommand{\cfg}{{\small \textsf{CFG}}}
\newcommand{\saturn}{{\small \textsf{SATURN}}}
\renewcommand{\dag}{{\small \textsf{DAG}}}

\newcounter{pppcount}
\newcommand{\pppreset}{\setcounter{pppcount}{0}}
\newcommand{\ppp}{\refstepcounter{pppcount}\mbox{{\color{blue}$\langle\arabic{pppcount}\rangle$}}}

\newcommand{\resource}{\mbox{\textsf r}}
\newcommand{\timing}{\mbox{\textsf t}}

\newcommand{\summarize}{\mbox{\cal S}}

\newcommand{\For}{\mbox{\textsf{for}}}
\newcommand{\Else}{\mbox{\textsf{else}}}
\newcommand{\If}{\mbox{\textsf{if}}}

\newcommand{\myiff}{\mbox{\texttt{iff}}}

\newcommand{\void}{\mbox{\textsf{void}}}
\newcommand{\assume}[1]{\mbox{\textsf{assume(#1)}}}
\newcommand{\assign}[2]{\mbox{\textsf{#1 := #2}}}

\newcommand{\exec}{\mbox{\textsf{exec}}}

\newcommand{\transition}[3]{#1~\xlongrightarrow[]{#3}~#2}
\newcommand{\shorttransition}[2]{#1~\xrightarrow[]{}~#2}
\newcommand{\trans}{\longrightarrow}
\newcommand{\shorttrans}{\rightarrow}
\newcommand{\translabel}[1]{\xlongrightarrow[]{#1}}
\newcommand{\loc}{\mbox{$\ell$}}
\newcommand{\locations}{\mbox{$\Sigma$}}

\newcommand{\transsystem}{\mbox{$\mathcal{P}$}}
\newcommand{\newtranssystem}{\mbox{$\mathcal{G}$}}

\newcommand{\symstate}{\mbox{$v$}}
\newcommand{\pci}[1]{\mbox{$\loc_{#1}$}}
\newcommand{\next}{\mbox{$\rightarrow$}}
\newcommand{\pc}{\mbox{\loc}}
\newcommand{\pcend}{\mbox{$\loc_{\textsf{end}}$}}
\newcommand{\pcerror}{\mbox{$\loc_{\textsf{error}}$}}
\newcommand{\pcstart}{\mbox{$\loc_{\textsf{start}}$}}
\newcommand{\pathcond}{\mbox{$\Pi$}}
\newcommand{\store}{\mbox{$s$}}
\newcommand{\pathcondbar}{\mbox{$\overline{\Pi}$}}
\newcommand{\storebar}{\mbox{$\overline{h}$}}
\newcommand{\symstatebar}{\mbox{$\overline{\symstate}$}}
\newcommand{\mapstatetoformula}[1]{\mbox{$\llbracket {#1} \rrbracket$}}

\newcommand{\sympath}{\mbox{$\pi$}}

\newcommand{\typevar}{\mbox{\emph{Vars}}}
\newcommand{\typesymvar}{\mbox{\emph{SymVars}}}
\newcommand{\typeop}{\mbox{\emph{Ops}}}
\newcommand{\typefo}{\mbox{\emph{FO}}}
\newcommand{\typeterms}{\mbox{\emph{Terms}}}
\newcommand{\typestate}{\mbox{\emph{States}}}
\newcommand{\typesymbstate}{\mbox{\emph{SymStates}}}
\newcommand{\typesympath}{\mbox{\emph{SymPaths}}}
\newcommand{\true}{\mbox{\frenchspacing \it true}}
\newcommand{\false}{\mbox{\frenchspacing \it false}}
\newcommand{\typebool}{\mbox{\emph{Bool}}}
\newcommand{\typeint}{\mbox{\emph{Int}}}
\newcommand{\typenat}{\mbox{\emph{Nat}}}
\newcommand{\typekeys}{\mbox{$\mathcal{K}$}}
\newcommand{\typevoid}{\mbox{\emph{Void}}}

\newcommand{\tick}{\mbox{\frenchspacing \it tick}}
\newcommand{\appr}{\mbox{\frenchspacing \it approx}}

\newcommand{\eval}[2]{\llbracket {#1} \rrbracket_{#2}}
\newcommand{\define}{\mbox{~$\triangleq$~}}

\newcommand{\unknown}{\mbox{\textsf{$\circ$}}}

\newcommand{\Intpsymbol}{\mbox{$\overline{\Psi}$}}
\newcommand{\InvariantFunc}{\mbox{\textsf{invariant}}}
\newcommand{\InvariantSym}{\mbox{$\mathcal{I}$}}
\newcommand{\ConflictSym}{\mbox{$\mathcal{C}$}}
\newcommand{\ContextSym}{\mbox{$\mathcal{O}$}}
\newcommand{\modifies}{\mbox{\textsf{\textsc{Modifies}}}}
\newcommand{\havoc}{\mbox{\textsf{\textsc{Havoc}}}}
\newcommand{\getvars}{\mbox{\textsf{var}}}
\newcommand{\getassrt}{\mbox{\textsf{getassrt}}}

\newcommand{\absdomain}{\mbox{$\mathcal{A}$}}
\newcommand{\answer}[1]{\theta_{#1}}
\newcommand{\Answer}[1]{\Sigma_{#1}}
\newcommand{\widening}{\mbox{$\nabla$}}
\newcommand{\narrowing}{\mbox{$\triangle$}}
\newcommand{\absfunc}{\mbox{$\alpha$}}
\newcommand{\concfunc}{\mbox{$\gamma$}}
\newcommand{\pre}{\mbox{\textsf{$\widehat{pre}$}}}
\newcommand{\post}{\mbox{\textsf{$\widehat{post}$}}}
\newcommand{\wfix}{\mbox{\textsf{wFix}}}
\newcommand{\fix}{\mbox{\textsf{Fix}}}

\newcommand{\state}{\mbox{$\ell$}}
\newcommand{\safety}{\mbox{$\psi$}}
\newcommand{\cons}{\mbox{$\phi$}}
\newcommand{\target}{\mbox{$\gamma$}}
\newcommand{\solutions}{\mbox{$\Gamma$}}
\newcommand{\id}[2]{\mbox{\textsf{Id}($#1, #2$)}}

\renewcommand{\If}{\mbox{\textbf{if}}}
\newcommand{\Endif}{\mbox{\textbf{endif}}}
\newcommand{\Return}{\mbox{\textbf{return}}}
\newcommand{\Then}{\mbox{\textbf{then}}}
\renewcommand{\Else}{\mbox{\textbf{else}}}
\newcommand{\Continue}{\mbox{\textbf{continue}}}
\newcommand{\Foreach}{\mbox{\textbf{foreach}}}
\newcommand{\While}{\mbox{\textbf{while}}}
\newcommand{\Do}{\mbox{\textbf{do}}}
\newcommand{\Until}{\mbox{\textbf{until}}}
\newcommand{\Done}{\mbox{\textbf{done}}}
\newcommand{\Goto}{\mbox{\textbf{goto}}}
\newcommand{\Let}{\mbox{\textbf{let}}}
\newcommand{\In}{\mbox{\textbf{in}}}
\newcommand{\Endfor}{\mbox{\textbf{endfor}}}
\newcommand{\Endwhile}{\mbox{\textbf{endwhile}}}

\newcommand{\authornote}[3]{
    {\fbox{\sc #1}:$\blacktriangleright$\textcolor{#2}{\small{#3}}$\blacktriangleleft$}%
}
\newcommand{\Vijay}[1]{\authornote{Vijay}{red}{#1}}


\textfloatsep 3mm plus 1mm minus 0mm
\dbltextfloatsep 6pt plus 2pt minus 2pt
%

\title{Incremental Quantitative Analysis on Dynamic Costs}


\authorinfo{Duc-Hiep Chu}
           {National University of Singapore}
           {hiepcd@comp.nus.edu.sg}
\authorinfo{Joxan Jaffar}
	{National University of Singapore}           
	{joxan@comp.nus.edu.sg}     
	
\authorinfo{Vijayaraghavan Murali}
	{Rice University}           
	{vijay@rice.edu}

\maketitle 




\begin{abstract}

In quantitative program analysis, values are assigned to execution traces  
to represent a quality measure.  Such analyses cover important
applications, e.g. resource usage.  Examining all traces is well known
to be intractable and therefore traditional algorithms reason over an
over-approximated set.  Typically, inaccuracy arises due to
inclusion of infeasible paths in this set.  Thus
\emph{path-sensitivity} is one cure.  However, there is another reason for
the inaccuracy: that the cost model, i.e., the way in which the analysis
of each trace is quantified, is \emph{dynamic}.  That is, the cost of
a trace is dependent on the context in which the trace is executed.
Thus the goal of accurate analysis, already challenged by
path-sensitivity, is now further challenged by \emph{context-sensitivity}.

In this paper, we address the problem of quantitative analysis defined
over a dynamic cost model.  
Our algorithm is an ``anytime'' algorithm: it generates
an answer quickly, but if the \emph{analysis resource budget} allows, 
it progressively produces better solutions via refinement
iterations.  The result of each iteration remains sound, but
importantly, must \emph{converge} to an exact analysis when given
an unlimited resource budget.  
In order to be scalable, our algorithm is designed to be \emph{incremental}.
We finally give evidence that a new level of practicality is achieved
by an evaluation on a realistic collection of benchmarks.



\end{abstract}


\ignore{
An idealized algorithm should be efficient, i.e., generating answers
quickly; but if the  \emph{resource budget} allows, should progressively
produce better solutions via a number of refinement iterations.
A pioneer in this direction is \cite{cerny13popl},
a CEGAR-based approach.
In this paper, we present an alternative approach for iterative refinements,
built on top of symbolic execution. We produce a 
scalable algorithm with several desirable properties and demonstrate it
with real programs on two kinds of analyses: a timing analysis and a
data flow analysis.  We show that in many cases,
our iterative method is in fact superior to both
Abstract Interpretation (AI) approaches as well as
algorithms designed to run continuously till an exact
analysis is found.
}

\ignore{
We consider the quantitative analysis of programs where execution traces are
assigned numerical values to represent a quality measure.  Such analyses
cover important applications particularly for resource usage.
An idealized algorithm should be efficient, i.e., generating answers
quickly; but if the  \emph{resource budget} allows, should progressively
produce better solutions via a number of refinement iterations.
A pioneer in this direction is \cite{cerny13popl},
a CEGAR-based approach.
In this paper, we present an alternative approach for iterative refinements,
built on top of symbolic execution. We produce a 
scalable algorithm with several desirable properties and demonstrate it
with real programs on two kinds of analyses: a timing analysis and a
data flow analysis.  We show that in many cases,
our iterative method is in fact superior to both
Abstract Interpretation (AI) approaches as well as
algorithms designed to run continuously till an exact
analysis is found.
}

\ignore{
  OLD STUFF BELOW

Program analysis has been dominated by Abstract Interpretation (AI),
owing to its scalability. AI is typically not (or only partially) path-sensitive
thus the obtained level of accuracy could be arbitrarily low.
Recently, there have been works on path-sensitive program analysis,
applied to domains where accuracy is critical.
However, they suffer from the path explosion problem and 
are not scalable in general.
In this paper, we present a general framework for program analysis that {\em incrementally} 
increases accuracy as it iterates. 
%


We start with a formulation of analysis with the branch-and-bound paradigm,
where an analysis result can be a lower or upper bound of the program's
behavior.  This duality allows for a specification
of precision in the overall analysis.
We then define an abstract representation of the program
which we can iteratively refine.
The iterations allows
for \emph{early termination} when a user-definable 
level of precision in the analysis has been extracted.
The critical performance factors are
(a) the incrementality of the refinement step where results from previous iterations
are persistent for future iterations,
(b) reuse of analysis from subproblems that have precise analysis,
and most importantly, (c) a concept of \emph{domination} which allows
a lower-bound analysis to prune away subproblems.

%
}




\section{Introduction}
\label{sec:intro}

\ignore{
 On the other hand, a
\emph{quantitative} objective assigns to every run of a program a real
value that represents a quality measure of the run.
}

In a \emph{qualitative} analysis of programs, such as testing, model
checking and verification, we assign to every execution trace of a
program a Boolean value: accept or reject. In
contrast, in \emph{quantitative} analysis, each trace is assigned
a quantity value or \emph{cost}, and the analysis estimates the collection of
such values into an overall quantity measure. Ideally one would like to compute
an optimal quantity measure in a given budget. Quantitative
analysis covers a wide range of important applications such as
Worst-Case Execution Time (WCET) analysis
(see \cite{WCETOverview00,WCETOverview08} for surveys), power
consumption
\cite{tiwari94iccad}, performance testing \cite{banerjee13rtss}, to name a few.   
Another class of applications involves detecting and quantifying 
the amount of information leakage. For example, this can be attempted via some
form of data flow analysis.


Quantitative program analysis has been so far dominated by 
some form of Abstract Interpretation (AI) where abstract  
properties are propagated through transitions induced by the program.  
(In WCET analysis, \cite{theiling00wcet}
proposes an efficient cache domain while
interval abstraction is used in \cite{TuBound}.)
Typical AI implementations are efficient and scalable; however, 
their precision could be arbitrarily low, and perhaps more
importantly, the level of (im)precision is \emph{unknown}.

Thus there is a great need to sometimes go beyond an efficient implementation of
AI.  Now a principal reason for the efficiency of AI is that it has little
consideration for \emph{path-sensitivity}, due to its abstract
reasoning.  Path sensitivity, on the other hand, faces the
challenge of the path explosion problem.  In fact, we can focus on a
sub-problem of this general problem: how to make sure
certain \emph{infeasible paths} do not distort the analysis result.
Addressing this sub-problem are many works that refine the process of
AI, perhaps the most notable are the CEGAR
\cite{clarke00cegar}  based approaches which
refine the abstract domain after having identified a so-called
``counterexample'' path as a
possible cause for distortion.

Dealing with path-sensitivity is, however, only half the story.
Another important cause of inaccuracy in analysis is due to the
fact the cost model, that is, the way in which the analysis of each trace is quantified, 
is \emph{dynamic}.   More specifically,
this means that the quantitative measure of a trace is dependent 
on the context in which the trace is executed.
Thus the goal of accurate analysis, already challenged by
path-sensitivity, is now further challenged by \emph{context-sensitivity}.

In summary, current analysis algorithms are inaccurate for two main reasons:
they include traces without consideration of \emph{feasibility},
and also include traces without consideration of \emph{optimality}.

The class of analysis problems which employ a dynamic cost model is significant.
We have mentioned two examples above.
The first is the class of resource analysis over low-level programs.
Here the dynamism arises from the underlying micro-architecture,
with the cache as the prominent example.
To see that the cost model is in fact dynamic is easy: running a trace
starting from a different initial cache configurations may clearly end up with
different results (for timing, or energy usage).
A different class is that of ``forward'' analyses where the cost of a trace
is intimately dependent on a prefix.  Forward data flow analysis
is an example, and this kind of analysis is clearly similar to many others,
e.g., points-to analysis.

In this paper we present an algorithm for accurate
quantitative analyses over a dynamic cost model that
 that makes the best use of a given budget.
The algorithm is based on abstract symbolic execution, exploring the
symbolic execution space while performing judicious abstraction in
order to achieve scalability.  Its main loop iterations
perform \emph{refinement} to the previous level of abstraction, so as
to enhance the accuracy of the analysis.  Its has two key features:
(a) It answers quickly in one iteration with a sound analysis,
and successive iterations can only improve the analysis.
Therefore it is an ``anytime'' algorithm \cite{boddy91aaai}.
More importantly, if the analysis resource budget is sufficiently large, then
it converges in the sense that it eventually produces
an \emph{exact} analysis.  
In other words, our algorithm is \emph{progressive}.
(b) It can compare its latest answer, using a lower bound
on the quality of the analysis, with a worst case estimate of any other
possible answer, i.e., an upper bound.  Therefore we also have the
important practical feature ``early termination'' when the current answer is
deemed good enough.

The main technical challenge we address is, as usual, scalability.
Each iteration, in its quest for more accuracy, embodies more detail
and thus, the level of detail grows exponentially.
Therefore, we have designed our algorithm to be \emph{incremental}.
This means that we require:

\begin{itemize}
\item
a \emph{persistent and compact} representation of the analysis on each iteration, and 
\item
an ability to \emph{reuse} (parts of) the analysis of previous
iterations as we refine.

\end{itemize}

\noindent
We achieve this by having an effective \emph{pruning} of the search space
by  using an established technique of \emph{reuse}
facilitated by the computation of interpolants and witness paths,
and maintaining lower and upper bounds on parts of the subspace,
and thus \emph{branch-and-bound} pruning is applicable.

Finally, In Section \ref{sec:experiments} we demonstrate our algorithm 
on the most prominent of quantitative analysis: WCET.
With realistic benchmarks, we show that the incremental 
iterations indeed produce precision gains progressively, and the final
analysis is always more precise than that obtained through AI.
Importantly, in many benchmarks, our algorithm terminates (i.e., producing
an exact analysis) \emph{faster} than the best custom algorithms
that are designed to pursue an exact analysis in one iteration.
Our experiments also show that our method can analyze programs that are known
to be particularly hard to analyze in the WCET community.

\ignore{
\todo{In one para, summarize the abstract and refine algorithm}
A key feature is that our algorithm is \emph{progressive}.
Our algorithm is also ``anytime'' in the sense that it is efficient enough
to generate answers quickly in one iteration, and 
if the \emph{resource budget} is sufficiently large, then our algorithm also 
\emph{converges} in the sense that it eventually produces an \emph{exact} analysis.
Finally, our algorithm can compare its latest answer (as a lower bound
of the cost quality) with a worst case estimate of any other
possible answer (an upper bound).  Therefore we also have the
important feature ``early termination'' when the current answer is
deemed good enough.

There they considered quantification
over a numerical value, and thus the work is generally targeted to
resource analysis.  
}

\ignore{
\bigskip
\todo{MOVE to Related ???}
We mention here the most related work \cite{cerny13popl}
which also performed quantitative analysis over a dynamic cost model,
and whose main loop iterations perform abstraction refinement
but this time in the CEGAR \cite{cegar} framework.
The key technical difference between this work and ours 
is that this work refines the abstract domain, while we refine 
the transition system.
More specifically, 
we iteratively refine the Control Flow Graph (CFG) with appropriate splitting. 
The relationship of our refinement step to \cite{cerny13popl}'s  
is akin to that of Abstract Conflict Driven Clause Learning (ACDCL) \cite{silva13popl} 
to traditional CEGAR in the context of program verification. 
A direct tradeoff is that we need to maintain a data structure called the 
\emph{hybrid symbolic execution tree} (HSET).
But the gain is potentially significant; we quote: ``ACDCL never
changes the domain, and this immutability is crucial for efficiency
(over CEGAR), because the implementations of the abstract domain and
transformers can be highly optimized'' \cite{silva13popl}.
In the end, the algorithm of \cite{cerny13popl} is not incremental;
the result of one iteration is not persistent and used in the next iteration,
and so the algorithm is not scalable.
We shall elaborate on the abovementioned related works in Section \ref{sec:related}.
}

\ignore{
Quantitative analysis is in fact well-suited for an 
``anytime algorithm'' \cite{boddy91aaai}, as realized in \cite{cerny13popl}.
An idealized algorithm should be efficient, i.e., generate answers quickly; but if the  
\emph{resource budget} allows, then it should progressively produce better answers
via a number of refinement iterations.
The result of each iteration should of course remain sound, but importantly, 
must (theoretically) \emph{converge} to the ``exact analysis'' when given infinite resource budget.

The most important work that pursues this ideal
is \cite{cerny13popl}.
There they proposed state-based and segment-based
abstraction schemes, coupled with algorithms for counterexample-guided
abstraction refinement (CEGAR), now extended for quantitative properties.  
The idea is that each iteration produces an 
analysis result together with an \emph{extremal} trace,
called the {\em ext-trace}, witnessing that result. The refinement
process then targets this ext-trace: either confirms or refutes it.
The reason for targeting an extremal trace is simple;
a refinement which does not refute this trace would 
not improve the analysis.
}




\ignore{
Unfortunately, this algorithm does not seem to scale to realistic programs for
the following reasons:

\begin{itemize}

\item[(1)] Being based on CEGAR, this algorithm suffers from \emph{expensive refinement steps}.
  It is now well known
  that a refined abstract domain is next applied globally
  even though the refinement was obtained from a specific execution trace.
  Thus for example, when considering a program {\tt if (b) S1 else S2}
  if a refined abstract domain is obtained from the consideration of \texttt{S1} alone, then
  this domain will also be subject to the analysis of \texttt{S2} even though the refinement may
  be irrelevant here.  Recall that once a refinement is made, it is not abstracted later.
  This fundamental issue of CEGAR is often referred to as [it] ``cannot recover from too-specific refinements''
  \cite{mcmillan10cav}.
   
\item[(2)] \cite{cerny13popl} does not possess \emph{linear progress}.
  By this we mean that an execution trace can be reconsidered repeatedly.
  Recall that the analysis phase distinguishes one ext-trace.
  The next step is to check if this trace indeed concrete.
  If not, a refinement is performed.
  Now in traditional CEGAR applied to verification, the distinguished trace
  is a possible \emph{counterexample} to verification.
  Checking if this trace is safe is tantamount to checking the \emph{feasibility}
  of a path, and if found to be infeasible, then refinement guarantees
  that in the next iteration, this trace will
  not be selected again.  In quantitative analysis, on the other hand,
  the ``counterexample'' trace may not be optimal for a reason
  \emph{different} from feasibility: that the current abstraction is not
  precise enough to report a sufficient quality.
  That is, even if the trace is feasible, it may not produce the analysis
  expected from it in reality due to the imprecision of the abstract domain used. 
  Now, once this domain is refined, a trace previous ruled as sub-optimal 
  could have become extremal,  and will have to be considered again. For this reason, abstraction
  refinement of \cite{cerny13popl} cannot guarantee that the same trace will not
  be reconsidered (unless, of course, it is infeasible).

   
\end{itemize}
}

%


\noindent

\section{Overview and Examples}
\label{sec:example}

\ignore{
Instead of iteratively refining the abstract domain
as in CEGAR, 
we refine the Control Flow Graph (CFG) with appropriate splitting. 
The relationship of our refinement step to \cite{cerny13popl}'s 
is like that of Abstract Conflict Driven Clause Learning (ACDCL) \cite{silva13popl} 
to traditional CEGAR \cite{clarke00cegar} in the context of verification. 
A direct tradeoff is that we need to maintain a data structure called the 
\emph{hybrid symbolic execution tree} (HSET).
But the gain is potentially significant; we quote: 
``ACDCL never changes the domain, 
and this immutability is crucial for efficiency (over CEGAR), 
because the implementations of the abstract domain and 
transformers can be highly optimized'' \cite{silva13popl}.
}

The conceptual core of our algorithm is centered on the \emph{symbolic
execution tree} (SET) of a program -- a tree representing all
possible symbolic paths.  Before proceeding, we first clarify that
in order to deal with a finite SET, we do not deal with unbounded loops. 
This is because we are performing a quantitative analysis, 
and in such an analysis, it is standard that there is an priori bounds on loops.
If we did not have this restriction, the analysis problem becomes
\emph{parametric}, and this is outside the scope of this paper.
For bounded loops, the general approach we use is to statically \emph{unroll} them.

In our setting, each (full) symbolic path in
the theoretical SET is interpreted using the most precise abstract
domain available.  Consequently, from the SET the ``exact'' analysis
can be extracted.  The SET is often too big to compute explicitly, we
instead compute a smaller \emph{hybrid SET} (HSET), a SET where some
subtrees of symbolic paths may be replaced by \emph{AI nodes}.  Each
node in a HSET is adorned with an analysis, which we shall call its
\emph{upper bound}.  Now an AI node is, intuitively, an
over-approximation of the analysis of the subtree it replaces but is
\emph{efficiently} obtained through abstract interpretation using some
coarse abstract domain.  Though an AI node is conceptually a single
leaf node in our HSET, we assume that as a by-product of its
(abstract) analysis, an AI node carries with it an \emph{extremal
  path} which displays the optimal analysis value over all paths from
the root to this AI node.  (Note again that since the analysis here
performed with abstraction, it is not necessary that the extremal path
is feasible nor optimal.)  Finally, if the subtree of a non AI node 
does not contain any AI node, then its upper bound will be \emph{exact}.
At this point, we say this bound is also the \emph{lower bound} of
the (analysis of the) subtree.

The main idea then is to define a refinement of a HSET, and this means
to choose an AI node to refine into a HSET, leaving all other nodes
unchanged.  Having chosen this node, we then use its extremal path in
order generate a symbolic execution path or ``spine'' eminating from
the node.  Along this path, we construct new AI nodes along each
branch deviating from the spine, and thus finally get a new HSET to
replace the chosen AI node.  Clearly the new HSET exhibits more
information because the spine exhbits exact symbolic execution, and
further, each new AI node deviating from the spine has a context emerging
from exact symbolic execution propagated along the spine up to the
deviation point.  Finally, how do we choose an AI node?
In the base case where the tree constains no AI nodes, 
its root node will indicate an exact analysis. 
In the general case, we now need to choose one AI node in the tree
to refine, that is, to replace it with a HSET which,
hopefully, will contain a more precise analyses than the AI node itself. 
The following choice, in conjunction with the use of the potential witness
paths, is what makes our algorithm goal-directed:
choose an AI node N which is has a \emph{maximal} upper bound.
(Not choosing this node means that its analysis will eventually
have to be refined later anyway.)


\ignore{
Otherwise, our symbolic execution would have driven a path --
what we call a ``spine'' -- through the HSET.  In other words,
an AI node in the HSET is replaced 
with a spine and new AI nodes along each branch deviating from the spine.  
Our new HSET exhibits more information than the previous one because of two reasons: 
(i) 
being inspected with the most precise abstract domain available,
the spine has been subjected to an exact
analysis, (ii) each new AI node 
can utilize the
precise information propagated along the spine up to the deviation point.
}

\begin{figure}[!t]
\begin{center}
\includegraphics[scale=0.5]{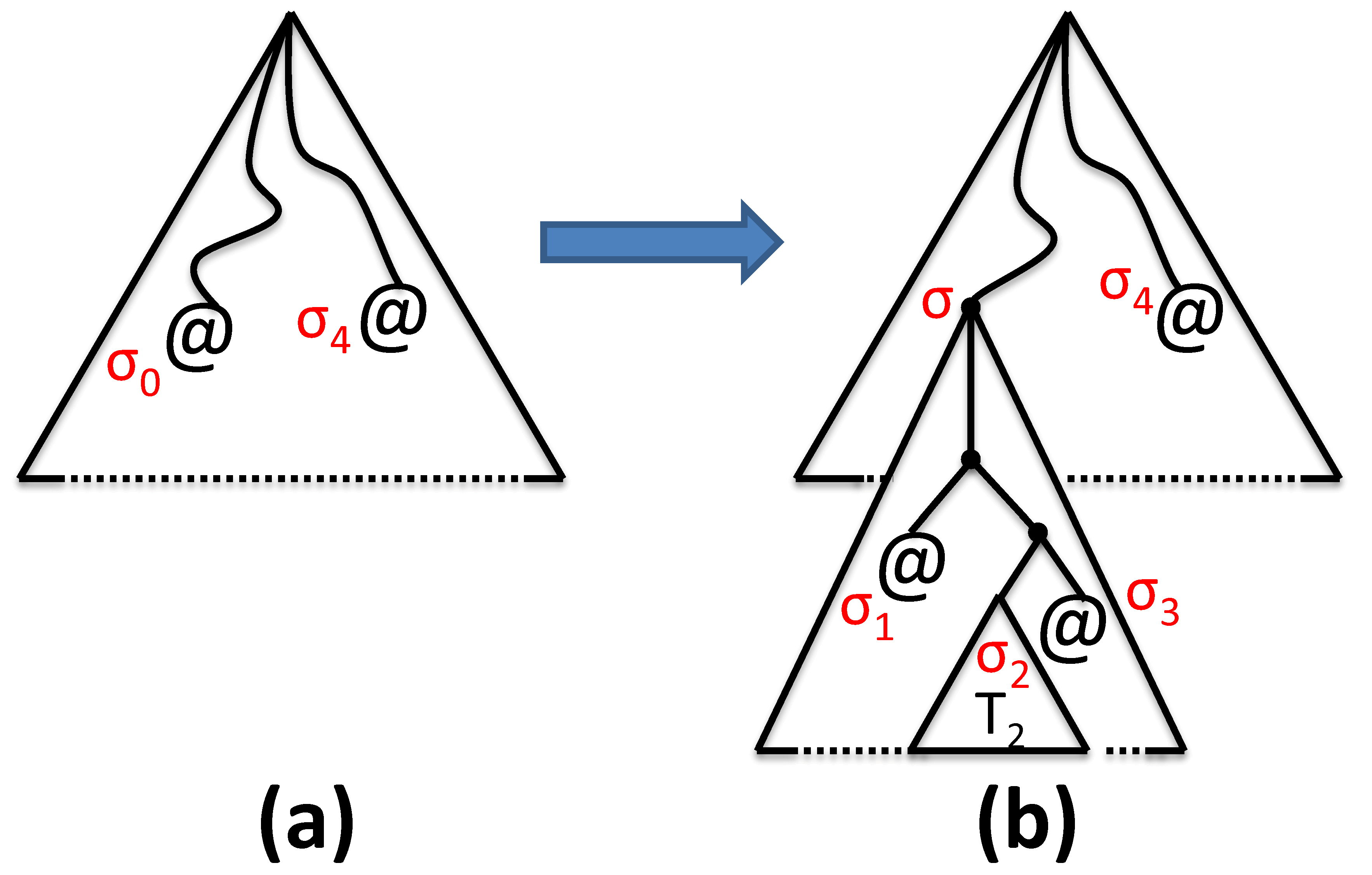}%
\end{center}
\caption{The Refinement Step}
\label{fig:abs-example}
\vspace{-2mm}
\end{figure}

\subsection{An Abstract Example}

We now walk through the HSET refinement process on an abstract example, 
in Figure \ref{fig:abs-example}(a).
The leftmost AI node @ with
upper bound $\sigma_0$ is refined into the subtree labelled with the
upper-bound analysis $\sigma$ in the second tree in Figure \ref{fig:abs-example}(b). Note that
this new subtree contains two AI nodes with upper bounds $\sigma_1$ and
$\sigma_3$. Note also that the subtree T2 does not contain any AI-nodes, and
so $\sigma_2$ is also an exact analysis.  We now detail this refinement. 

Next consider the AI node labelled $\sigma_0$ in the first tree.
Though this is a single node, we assume that the AI algorithm that gave rise
to its analysis $\sigma_0$ also gave information about its extremal path, say $p$.
Suppose this path, when expanded out from that single node, would go through
the subtree $T_2$ indicated in the second tree.
We then construct the subtree starting from $\sigma_0$ by first constructing 
the edges and nodes as a symbolic path p.  As
each edge and destination node is constructed from a branching source
node, we also construct a new edge and destination node corresponding
to the alternative of the branch. For this second destination node,
which is not in the path p, we now construct a new AI node. At the end
of this process, we would have constructed the subtree which has a
spine corresponding to the path p, and along the spine, we have
constructed a number of AI nodes (two, in this example, labelled $\sigma_1$ and $\sigma_3$).

See Figure Figure \ref{fig:abs-example}(b) 
and once again focus on the subtree labelled $\sigma$, and
where the spine is some path that includes $\sigma_2$.
There are two possible benefits of
this refinement step. One is that this sub-HSET $\sigma$ is more precise than
the original analysis $\sigma_0$ because the join of $\sigma_1, \sigma_2$ and $\sigma_3$ is more
precise than $\sigma_0$.
Another benefit is when $\sigma_2$, which is an exact analysis, can be
used to dominate any other analysis. For example, if the upper bound of $\sigma_4$
is less than the analysis of $\sigma_2$, then the entire subtree
at $\sigma_4$ can be pruned from further consideration.

We remark here in the refinement step, each of the newly generated AI
nodes require an (abstract) analysis, and although these analyses
are efficient, there is the issue that the \emph{number} of analyses could be
as long as the path p. However, an important feature is that in the
several invocations of abstract analysis performed here over the
several AI nodes, and because the employed abstract domain is coarse, 
the analysis of each of these is often produces the
the \emph{same} results, and hence can be cached and need not be redone. We
will argue and demonstrate this important feature in detail later.


\ignore{
The first advantage we have is that we do not suffer from the problems (1) and (2).
A reason for this is that we operate using symbolic execution,
which is in some sense the best abstract domain (since it can be 
made to track the logic of the execution flow precisely),
and refine the
Control Flow Graph (CFGs) with appropriate splitting.
In contrast, \cite{cerny13popl} refines
the abstract domain, thus affecting  \emph{both} CFGs and the quantitative 
properties (e.g., timing over abstract caches) at the end of each iteration.
In other words, the relationship of our refinement step to \cite{cerny13popl} 
in the context of quantitative analysis
is like that of Abstract Conflict Driven Clause Learning (ACDCL) \cite{silva13popl} 
to traditional CEGAR \cite{clarke00cegar} in the context of verification.
We quote: ``ACDCL never changes the domain, 
and this immutability is crucial for efficiency (over CEGAR), 
because the implementations of the abstract domain and 
transformers can be highly optimized'' \cite{silva13popl}.
}


\subsection{A Motivating Example: Feasibility}

\ignore{
Consider here a trivial class of programs which contain only assignment
statements of the form $\tick ~+\!\!\!= \kappa$ where $\kappa$ is a
positive number, and consider only non-nested if-then-else statements
with unspecified guards $b_i$ which do not depend on the variable $\tick$.
The example analysis is an abstraction of the well-known worst-case
execution time (WCET) analysis, and in our simple setting, the analysis 
formulas are simply bounds on \tick{}, and the final analysis is to
determine the (upper) bound of $tick{}$ at the end.
}

\begin{figure}[!t]
 \begin{center}
\stuff{
\pp{0} $\tick = 0$ \\
\pp{1} if ($b_1$) $\tick +\!\!= 3$ \\
\pp{2} if ($b_2$) $\tick +\!\!= 2$ \\
\pp{3} if ($b_3$) $\tick +\!\!= 1$
}
\\
\hspace*{-3mm}\includegraphics[scale=0.8]{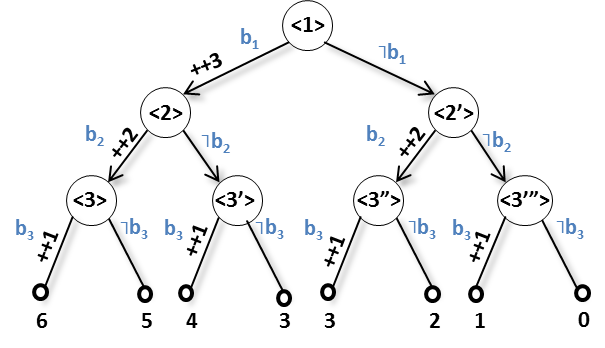}
 \end{center}
\caption{Example Program and its Symbolic Execution tree}
\label{fig2}
\vspace{-4mm}
\end{figure}

Consider the program and its SET in Figure~\ref{fig2}
and the WCET problem at hand is to determine the upper bound of \tick{}.
Assuming that any boolean combination of the unspecified guards $b_i$ is satisfiable.
Then clearly the WCET is $6$, obtained from the leftmost path.

\ignore{
Before we proceed, 
note that in the general case where not all combinations of guards
are satisfiable (as we will consider in this Section), 
and remaining within our trivial programming language,
the problem to find the WCET is NP-hard \cite{jaffar08aaai}.
}

To demonstrate reuse, assume that $\neg b_1 \wedge b_2 \wedge b_3$ is satisfiable, 
and that we already have an exact analysis, $\tick = 3$ of the right subtree marked \pp{2'}.
We now can produce an exact analysis for the left subtree marked
\pp{2} without having to traverse it.  To do this, we take the longest path in the
right subtree which gave rise to the analysis, i.e., the witness path,
and this is the leftmost path under \pp{2'}.  Call this path $p_1$.
We now \emph{replay} this path in the left subtree, getting the
leftmost path starting from the root.  Call this path $p_2$.  Now the
idea is that the analysis of $p_2$ is computed from the analysis of $p_1$,
which is $3$.  However, since the prefix of $p_1$ from the root to
node \pp{2'}, which increments \tick{} by zero, differs from the prefix
of $p_2$ from the root to node \pp{2}, which increments \tick{} by $3$,
we must adjust for this and now declare that the exact analysis of
node \pp{2} is $\tick = 6$.   In other words, we assumed that the
longest increment of \tick{} from node \pp{2} downwards is the same
as that from node \pp{2'}, which is $3$.  But since the prefix of
node \pp{2} is 3 more than the prefix of node \pp{2'}, we add
a further $3$ to obtain the final value $6$.

There are two further points to note about reuse.

\begin{itemize}
\item
If $b_1 \wedge b_2 \wedge b_3$ (ie. the leftmost path) were unsatisfiable,
reuse is in fact still \emph{sound}, when we declare that the analysis of node \pp{2}
is $6$.  But may be \emph{imprecise}.
To prevent imprecision, we check that the path under node \pp{2}
that corresponds to the ``witness'' is feasible.
\item
Now suppose $\neg b_1 \wedge b_2 \wedge b_3$ (leftmost path in the right subtree) is unsatisfiable 
but $b_1 \wedge b_2 \wedge b_3$ (leftmost path in the left subtree) is satisfiable.
Now it is \emph{unsound} to reuse the exact analysis of node \pp{2'} 
(which now is different from $3$) in the analysis of node \pp{2}.  
In previous implementations of reuse, 
e.g.: \cite{jaffar08aaai,jaffar12sas,chu11emsoft}, the exact analysis would be accompanied by an
\emph{interpolant} which would ensure that the reuse can soundly take place.

\ignore{
In a setting more general than the WCET example in this Section, we would
further need to check that the subtree to which reuse is considered
satisfies not just one or more feasible witness paths, but that the optimality
of these witness paths carries over from the source analysis.
}

\end{itemize}

\noindent
Our algorithm provides bounds for \tick\ in each node.
For example, an upper-bound analysis for the left subtree in
Figure~\ref{fig2}, labelled \pp{2}, is $\tick \leq 6$.
This subtree also can have a \emph{lower-bound analysis} of a nonnegative number
less than or equal to $6$; 
We also can have a lower bound.
for example, if the path proceeding to the left successor of \pp{3'} was feasible, 
$4 \leq \tick$ would be a lower bound.
If however we did not care to check the feasibility of any path going
through \pp{2}, then we could quickly estimate that $3$ is a lower bound
(by choosing only rightmost branches that do not add to \tick).
Note that there may not actually be a real execution path resulting in $\tick = 3$.
Note also that lower bounds whose values are too low
(e.g., $0 \leq \tick$) are not very useful.

\ignore{
An \emph{analysis} is defined over an ``abstract domain'' in the traditional manner 
\cite{cousot79ai}, and it provides an \emph{analysis formula} 
over a \emph{set of paths}.
Such a set is usually that which is associated by a symbolic state within some SET.
For example, where the analysis is about the upper bound
of the variable \tick, 
an \emph{upper-bound analysis} formula for the left subtree in
Figure~\ref{fig2}(a), labelled (\pp{2}, tick), is $tick \leq 245$,
When the lower and upper bounds coincide, we say the analysis is \emph{exact}.
}

\begin{figure}[!t]
\begin{center}
\includegraphics[scale=0.65]{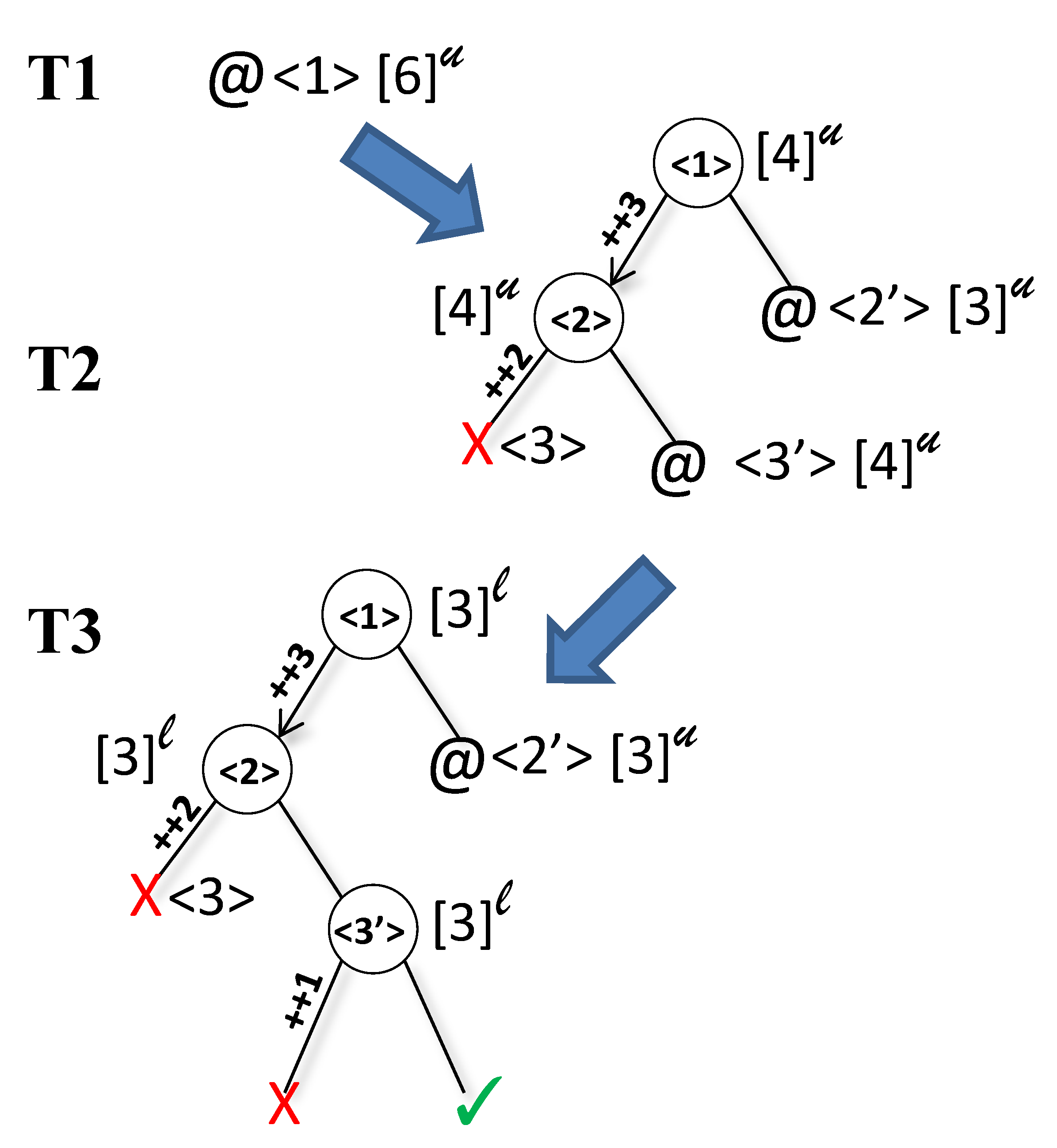}%
\end{center}
\caption{Detailed Refinement Step}
\label{fig3}
\vspace{-3mm}
\end{figure}

We now proceed to analyze the program incrementally.
See Figure~\ref{fig3} where ``@'' denotes an AI node, the $l$ and $u$ superscripts denote
lower and upper bounds respectively, and the $T_i$s represent the HSET we
construct in each iteration. We start with a single AI node at $T_1$ representing
an (abstract) analysis of the program starting at the beginning.
We could have used traditional abstract interpretation (AI)
which over-approximates the set of paths in the SET in order to limit
consideration to a small number of abstract states (typically, one state
per program point).  Thus AI analyzers are typically very efficient.
We then quickly, because the analyzer is path-insensitive,
determine a (trivial) lower bound of $0$ and an upper bound
of $6$.  Furthermore, the analyzer indicates that the leftmost
path is a \emph{witness} path, i.e., if it were feasible,
then it would indicate the true WCET.
In Figure~\ref{fig3}, we show only upper bounds when the lower bound is trivial.

Next we refine the single AI node $T_1$ into the HSET $T_2$
which now contains new nodes, amongst them two AI nodes
at \pp{2'} and \pp{3'}.
Using abstract interpretation, note that former has an upper bound of $3$,
while the latter has an upper bound of $4$.
We assume that the constraint $b_1 \wedge b_2$ is unsatisfiable,
and so the leftmost path in Figure~\ref{fig2} is in fact infeasible
(at just before program point \pp{3}). 
Now since node \pp{3'} has a bound $4$, this is inherited by
the parent node \pp{2}.  Finally, the root node \pp{1}
inherits the larger of the bounds of its successors, which are $3$ and $4$,
and so we obtain a final bound of $4$.
Now since $T_2$ contains AI nodes which contribute to this answer,
this analysis is not confirmed to be exact.

Finally we deal with the two remaining AI nodes in $T_2$, and choose
one of them to refine.  We choose the node \pp{3'} over \pp{2'}
because its upper bound is higher.  The intuition is this: if we instead chose
to refine the AI node with the smaller bound, the other AI node will
still need to refined in the future.  If, as we will show next,
we choose the AI node \pp{3'} with the higher bound, there is a chance that 
the remaining AI node can be dominated.  We now obtain $T_3$ by refining
this AI node.

This refinement produced two successors, and by assuming that the constraint
$b_1 \wedge \neg b_2 \wedge b_3$ is unsatisfiable, we have that the left
subtree of node \pp{3'} is an infeasible path.  The right subtree is a terminal
node, and so for the first time, we can declare that, since both subtrees of \pp{3'}
have no AI-nodes, \pp{3'} has a \emph{lower} bound\footnote{In fact the
upper bound of \pp{3'} is also 3, i.e., it has an exact analysis} of $3$.
The most interesting step now can be taken: the analysis here \emph{dominates}
the analysis at the one remaining AI node at \pp{2'}.
Note that the set of paths represented by \pp{2'} is nearly half of
all the paths.
By pruning away this subtree, we now have that the entire tree has no more AI nodes,
and we can now declare that the root node has an exact analysis of $3$.

\subsection{A Motivating Example: Optimality}

\begin{figure*}[!t]
\begin{center}
\subfigure[CFG]{
\includegraphics[width=0.2\textwidth]{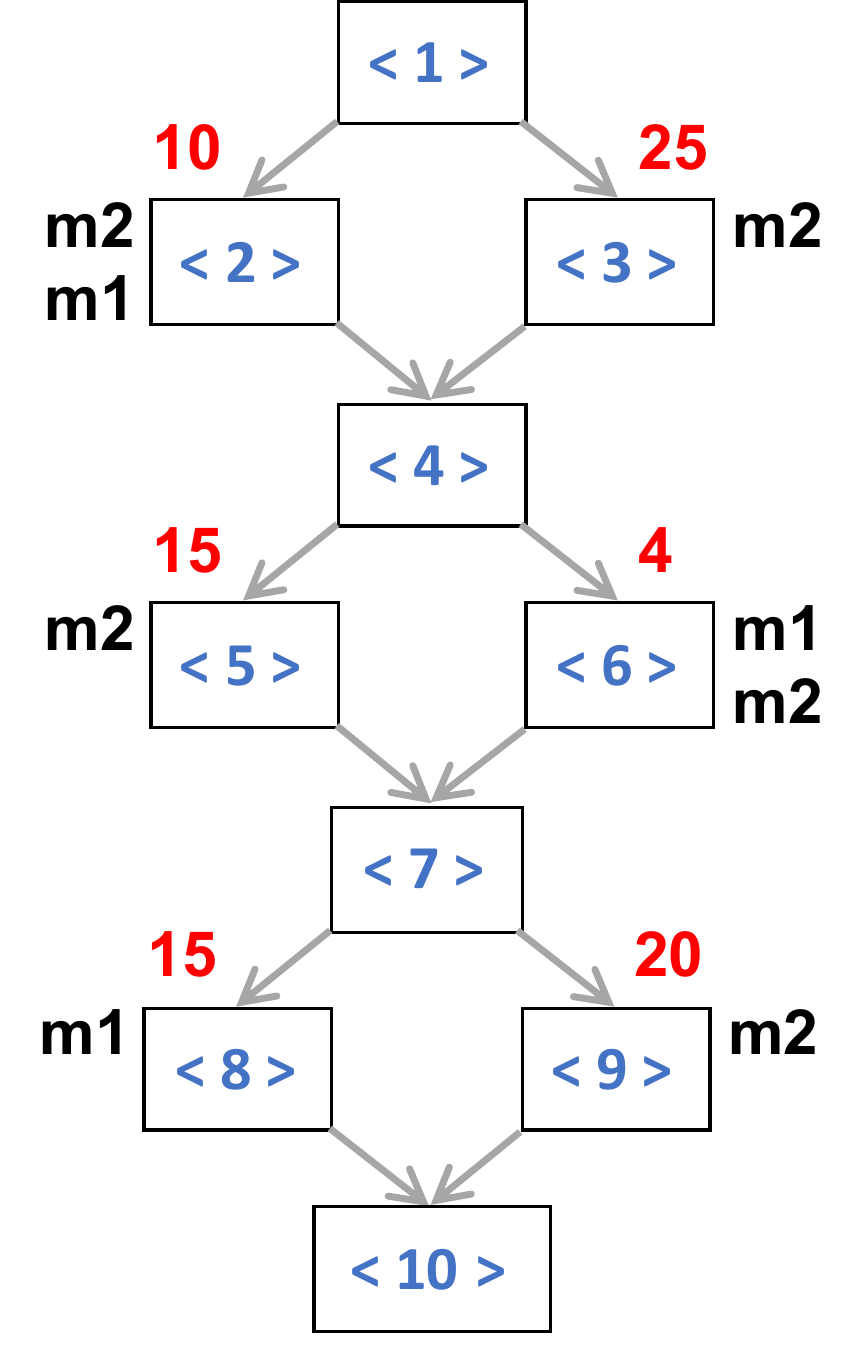}%
\label{fig:example:cfg}
}
\hspace{8mm}
\subfigure[Detailed Refinement Step]{
\includegraphics[width=0.55\textwidth]{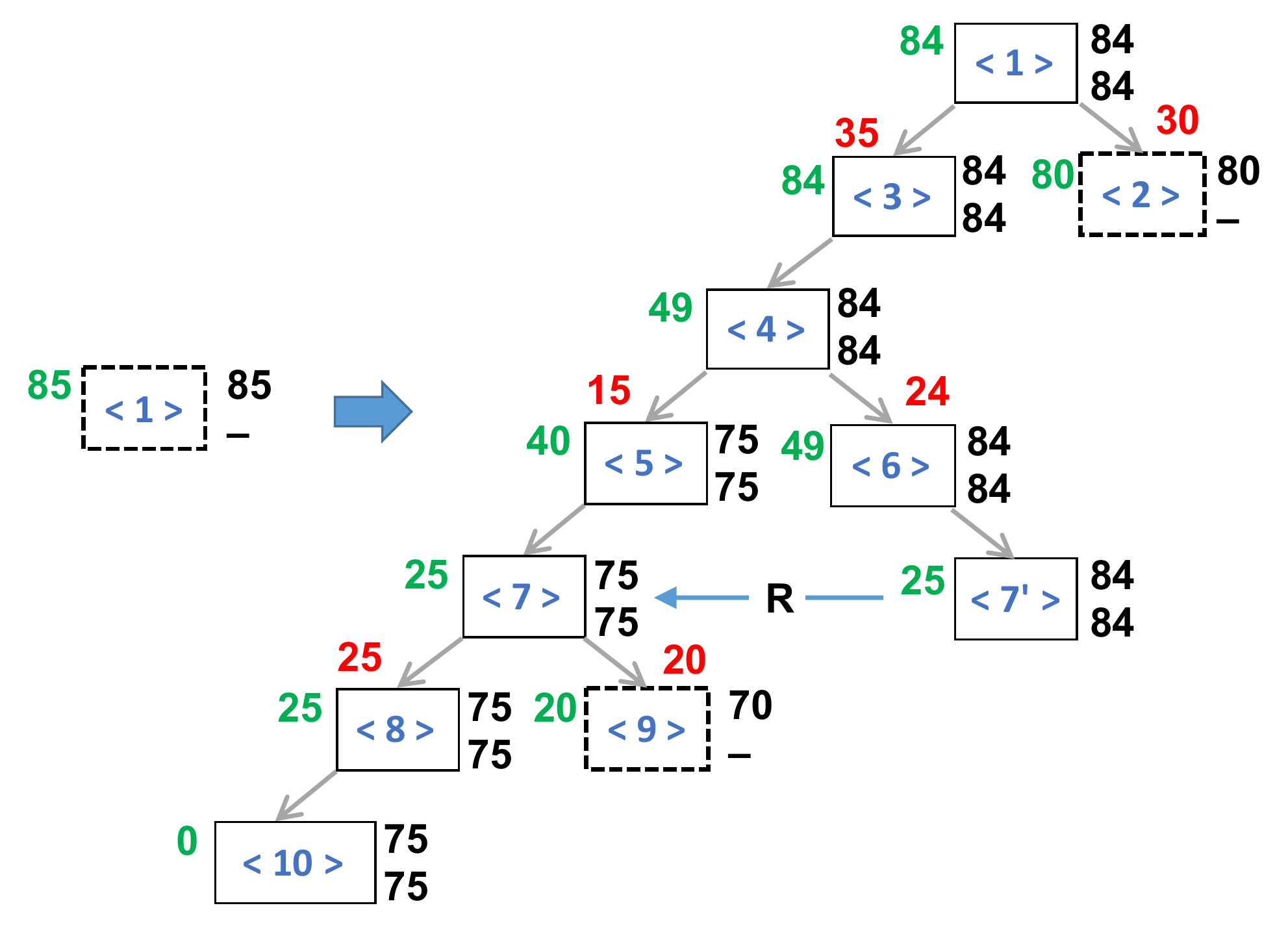}%
\label{fig:example:refine}
}
\caption{Example: WCET Analysis with Cache. 
The list of pair in (b) is just to simplify the presentation.}
\label{fig:example}
\end{center}
\vspace{-4mm}
\end{figure*}

We now consider an example with a dynamic cost model.
In particular, consider WCET analysis of the program, whose Control Flow Graph (CFG) is 
shown in Fig.~\ref{fig:example:cfg}. Each node -- rectangular box -- represents a basic block.
In the basic blocks, $\pp{1}, \pp{2}, \dots \pp{10}$ denote
the program points. While the timings of basic blocks
$\pp{1}, \pp{4}, \pp{7}, \pp{10}$ are always $0$,
other basic blocks are abstracted by the static timing of the
instructions in cycles, denoted by a non-negative integer 
(placed above each node, in red),  and a sequence of memory accesses $m_i$,
of which the timing depends on the cache configuration at the time of access.
(In the beginning the cache is empty.)

For simplicity, we assume: (1) direct-mapped cache;
(2) $m_1$ and $m_2$ map to the same cache location, i.e., they conflict;
(3) a cache miss costs $10$ cycles, while a cache hit costs $0$. 
Given the setting, it is clear that the timing of a single symbolic path can be
precisely determined. Thus if we exhaustively enumerate the symbolic
execution tree, exact analysis can be achieved. However, such approach
is \emph{prohibitive} in practice.

We now assume that the base Abstract Interpretation (AI) 
used is the must-analysis abstract cache proposed by \cite{theiling00wcet}.
To distinguish our approach from a large body of work
in program verification, we assume that all the symbolic paths
in the program are feasible. In other words, simply refining
on the (in)feasibility of paths \emph{will not improve} the analysis precision.

\vspace{1mm}
\noindent
{\bf Phase 1:} we start by invoking AI at $\pp{1}$.
Note that cache merging is perform at every join point.
While we can determine precisely that the accesses in $\pp{2}$, likewise
the access in $\pp{3}$, are all misses, the merge at $\pp{4}$
keeps neither $m_1$ nor $m_2$ in the must-cache. Similarly,
we can determine the accesses in $\pp{5}$ and $\pp{6}$ as all misses.
However, going through either $\pp{5}$ or $\pp{6}$, both paths end
up with $m_2$ in the cache, thus the merge at $\pp{7}$ keeps $m_2$.
Following up, the access to $m_1$ at $\pp{8}$ is a miss, while
the access to $m_2$ at $\pp{9}$ is a hit. 

In summary, the AI algorithm can give us an analysis
for each program point, summarizing the estimate of 
the \wcet{} from that point to the end of the program:

 ($\pp{10}$, 0), ($\pp{9}$, 20), ($\pp{8}$, 25), ($\pp{7}$, 25), ($\pp{6}$, 49), \\
 \indent
 ($\pp{5}$, 50),  ($\pp{4}$, 50), ($\pp{3}$, 85), ($\pp{2}$,80),  ($\pp{1}$, 85),

\noindent
where (B,T) means that T is the estimated worst-case timing from B 
to the end of the program. In other words,
invoking AI at $\pp{1}$, we achieve the analysis of 85 (= 35 + 25 + 25),
and the extremal trace witnessing that analysis is:
\vspace{-1mm}

\begin{center}
$\pp{1} \rightarrow \pp{3} \rightarrow \pp{4} \rightarrow \pp{5}
\rightarrow \pp{7} \rightarrow \pp{8} \rightarrow \pp{10}$.
\end{center}

\vspace{-1mm}
\noindent
{\bf Phase 2:} At the end of phase 1, our HSET
contains only one abstract node as in Fig.~\ref{fig:example:refine}.
We denote abstract nodes using dashed boxes.
Every node in our HSET, when applicable, will be 
annotated with 3 pieces of information:

\begin{enumerate}
\itemsep -3pt
\item the current estimate of  the \wcet{} from the node 
to the end of the program -- on the left, in green color;
\item the aggregated lower bound analysis of all the paths through
this node -- on the right and below; and
\item the aggregated upper bound analysis of all the paths passing through
this node -- on the right and above;
\end{enumerate}

\noindent
We proceed refining by first building a ``spine'' targeting
the extremal trace identified in the previous phase.
This (first) spine is shown as the left most path in Fig.~\ref{fig:example:refine}.
Note that at the end of the path, the analysis of $\pp{10}$
is exact, but the access $m_2$ at $\pp{5}$ has now resolved
to be a hit (instead of a miss), the annotation at $\pp{10}$ is 
therefore $0, [75]^{l}, [75]^{u}$ as shown. Similarly,
the annotation at $\pp{8}$ is $25, [75]^{l}, [75]^{u}$.
Note that the execution time of each block has been updated in
Fig.~\ref{fig:example:refine} considering the context in which
it is executed.

Now, we need to deal with analysis of the sibling node,
at program point $\pp{9}$. However, it is easy to see
that if we use the coarse estimate returned from the previous phase
for $\pp{9}$, domination happens, because the upper bound analysis
of all paths going through that node is just 70 ($= 35 + 15 + {\bf 20}$).
Correspondingly, the annotation for this node is $20, -, [70]^{u}$. 

Propagating back, we see that along the spine,
the analysis at $\pp{7}$ and $\pp{5}$ are now exact.
We now consider the sibling of $\pp{5}$, which is $\pp{6}$. 
Using the coarse estimate for $\pp{6}$ from the previous phase,
domination does not happen (because $35 + {\bf 49} > 75$).
We do not invoke a new AI analysis from here, but proceed until the 
next join point. There are two reasons:

\begin{itemize}
\itemsep -3pt
\item reuse of an exact analysis might be possible, as we will see soon. In this case, we achieve precise 
analysis for the node, while also avoid a call to the base AI analysis.
\item we can propagate information precisely till the join point (this is cheap),
so that the call to AI might return better analysis than what has been achieved
in the previous phases.
\end{itemize}

\noindent
So we proceed to the node $\pp{7}'$ in the figure. Note that
at $\pp{7}'$ we share the same cache context encountered before
in $\pp{7}$, i.e., $m_2$ is present in the cache. Also note that since
there are no infeasible paths, the {\em interpolant} \cite{chu11emsoft}
stored at $\pp{7}$ is simply $true$. Thus we can trivially reuse
the exact analysis of $\pp{7}$ at $\pp{7}'$.
The annotation at $\pp{7}'$ is then $25, [84]^{l}, [84]^{u}$.
The annotations for $\pp{6}$, $\pp{4}$, $\pp{3}$ can be computed from
this by propagating it back.

Fast forwarding, for $\pp{2}$, using the previous estimate, domination happens,
we then end up with the ``exact'' analysis for the whole program, 
as the annotation of $\pp{1}$ is $84, [84]^{l}, [84]^{u}$.

\ignore{
In summary, with a simple example, we have demonstrated:
(1) with dynamic costs (due to cache timing), other than feasibility, 
we need to consider the \emph{optimality} aspect of a trace for refinement;
(2) how the two concepts, namely \emph{reuse} and \emph{domination},
work in synergy to speedup the convergence of our algorithm to a precise -- 
and particularly in this example, exact --
analysis result.
}

\subsection{Discussion on Scalability}

We have already mentioned our algorithm is ``anytime'', and
further that it is progressive and consequently, the
algorithm converges in the sense that it eventually produces an
\emph{exact} analysis.  
The main reason for this that an execution
path is never be considered twice and so the search space is
monotonically strictly decreasing.  But of course this is not enough to
attain scalability.

As mentioned above, for scalability,
it is critical an algorithm which performs iterations
that are progressively more expensive to be incremental.  
That is, the work done in previous iterations must be both
(a) persistent and compact, (b) directly useful to mitigate the
cost of the next iteration.  We now overview how our algorithm
addresses these criteria by some form of pruning of the search space.

\ignore{ 642
We conclude here by noting that our algorithm has three desirable features:

\begin{itemize}
\vspace{-1mm}
\item[(1)] It has \emph{linear progress}:
  By this we mean that an execution trace can never be considered twice.
  In contrast, \cite{cerny13popl} does not possess this feature.
\item[(2)] It is \emph{incremental}:
  The results of the present iteration are persistent, 
  and can be (re-)used in the next iteration.
  As will be detailed later, information obtained
  from the current HSET is directly applicable when we deal with the next (refined) HSET.
\item[(3)] It maintains both a \emph{lower and upper bound} on the analysis:
This brings us in line with the optimization community, where it is standard to combine
a feasible/real solution, representing a ``lower-bound'' solution, 
with an ``upper-bound'' solution obtained by an efficient 
but not necessarily precise method.  This lower/upper interval 
provides for a termination criteria ahead of the budget exhaustion, 
so that we may have 
``early termination'' 
when the result is already within a satisfactory range.
\vspace{-1mm}
\end{itemize}

\noindent
To elaborate on (2), the follow-up question is 
whether incrementality ultimately translates into 
scalability on realistic programs.
Note that our refinement step can increase the number of
nodes in the HSET.   To address this, we employ:

\begin{itemize}
\vspace{-1mm}
\item \emph{re-use} of the analysis of a HSET node:
Suppose we have an exact analysis $E$ for a subtree rooted at a node
$\symstate$. In this scenario, we exploit $E$ to compute another exact analysis
$E'$ for another (yet unexplored) node associated with the same program point as $\symstate$.
\ignore{
In general, the condition for such a reuse as well as the precise
definition of the mapping from $E$ to $E'$, is quite involved
because it depends on the kind of analysis in question.
But in specific instances, this is easily done.
We thus omit a full description here but instead 
refer to \cite{jaffar08aaai,chu11emsoft,jaffar12sas},
and use an example of reuse in Section \ref{sec:example}.
}

\item \emph{domination} of a HSET node:
  By maintaining separate upper bound for each HSET node, we are able to \emph{prune}
  any HSET node whose potential analysis is already covered by the current lower bound.
  In short, this is a realization of the well-known \emph{branch-and-bound} strategy.
\end{itemize}

MUMBLE

As a result of the above, we are able to produce an efficient implementation.
We now outline some key reasons for the scalability of our method.
} 

\ignore{
Once again, we start with a HSET $T$ and now 
consider the refinement of node $N$ in $T$ into a new tree $T_1$, 
to be replaced in-situ as a subtree in $T$.   
We start with the requirement that $T_1$ has 
a nontrivial lower-bound analysis, or preferably, 
an exact analysis.
}

\vspace*{3mm}
\noindent
{\sc Reuse of Abstract Analyses:} \\
In the refinement of a node $\symstate$ to produce a spine path of length $n$,
we generally produce $n - 1$ new AI nodes attached to the spine.
However, it is typical in AI implementations (which was used 
on the node $\symstate$) to have computed analysis for all program points
that are reachable from $\symstate$'s program point (via the CFG),
and not just for that of the root $\symstate$.  Therefore much of the analyses
required for the new AI nodes are typically already at hand.

\ignore{
\vspace*{3mm}
\noindent
{\sc Scenario 2 (Infeasibility)}: \\
By exposing the spine, we are extending the path constraint that
is subject to satisfiability testing, and therefore increasing the
likelihood of discovering an infeasible path.
In this scenario, we exploit the situation where one of the successors of the node
being refined is unsatisfiable.  
This means that an entire subtree of symbolic paths,
which previously had been included in the AI node $\symstate$,
now has been removed.
}

\vspace*{3mm}
\noindent
{\sc Reuse of Exact Analyses:} \\
Here we use a computed exact analysis of one subtree to derive an exact 
analysis of another subtree.
Suppose we have an exact analysis $E$ for a subtree rooted at
$\symstate$. 
%
In this scenario, we exploit $E$ to compute another exact analysis
$E'$ for
another (yet unexplored) node associated with the same program point as $\symstate$.
In general, the \emph{witness condition} for such a reuse (here we are talking about real
witnesses for the exact analysis), as well as the precise
definition of the mapping from $E$ to $E'$, is quite involved
because it depends on the kind of analysis in question.
But in specific instances, this is easily done.
We thus omit a full description here but instead 
refer to \cite{jaffar08aaai,chu11emsoft,jaffar12sas},
and use an example of reuse in Section \ref{sec:example}.

\ignore{
and that it has a collection of witness paths.
(This time these are real witnesses, and not potential witnesses.)
Suppose for some other node $N_2$ anywhere in 
in the HSET, the program states of $T_1$ and $T_2$ represents 
the same program point\footnote{
We say such states are ``similar''.},
We now apply an essential concept \emph{reuse}.  
we need to check two things:
\begin{itemize}
\item whether the set of \emph{feasible paths} starting from $N_2$ is a subset of those starting
of $N_1$, and
\item whether the witness paths of $T_1$ are feasible when rooted at $N_2$. 
More precisely, take such a path of $T_1$, replace the prefix path that leads
to $N_1$ with the prefix path $p$ that leads to $N_2$, and then conjoin the suffix path
from $N_1$ to $N_2$
\end{itemize}
If these conditions are met, we can compute the exact WCET for $N_2$ as follows.
Suppose the contexts of $N_1$ and $N_2$ produce increments of $\tick$ of $n_1$ and
$n_2$ respectively.  Then we have that $n - n_1 + n_2$ is the exact bound for $N_2$.
We remark here that the first condition is required for soundness,
that is, if indeed $N_2$ contained a path which was infeasible in $N_1$,
the analysis can be wrong.  The second condition, on the other hand,
is required for \emph{precision}.  If we were not seeking an exact analysis,
we could omit this second condition.  This basic algorithm was first used
in \cite{jaffar08aaai} for ``longest path'' analysis such as in this
subsection; subsequent practical implementations were \cite{chu11emsoft} 
for general WCET analysis and \cite{jaffar12sas} variable-dependency analysis
for the purpose of program slicing.  
}

\vspace*{3mm}
\noindent
{\sc Domination:} \\
Suppose we have a nontrivial lower-bound analysis, say for node $\symstate$. 
Now we can in fact prune \emph{all} subtrees 
which are dominated by $\symstate$.
Note that domination does not require that the two entities
involved represent the same program point, in contrast to reusing.
In other words, \emph{any} node/subtree can dominate any other.
Another difference between reuse and domination is that both parties
involved in a reuse contribute an analysis; it is just that we have a quick
way to compute one of them from the other.  Domination however
means that we can simply ignore the \emph{dominated} party.

\vspace*{3mm}
In the end, during the refinement process,
the effects of reuse and domination serve to prune the search space,
while the increasing level of exact analyses of subspaces.
This produces more lower-bound analyses, and these, in turn, 
produce better upper bounds, and this in turn,
creates further opportunities for reuse and domination.
This cycle of mutual benefits is the key to the scalability
of our algorithm.

\section{Hybrid Symbolic Execution with Interpolation}
\label{sec:prelim}
\label{sec:preliminaries}

Here we provide the formalities required for our algorithm.
In particular, we cover the needed aspects of symbolic execution, abstract
interpretation and interpolation.  We also highlight some important assumptions.

\vspace{2mm}
\noindent \textbf{Syntax}. We restrict our presentation to a simple
imperative programming language 
where all basic operations are either assignments or assume
operations, and the domain of all variables is the integers. The set of
all program variables is denoted by \typevar.  An
\emph{assignment} \assign{x}{e} corresponds to assigning the evaluation
of the expression \textsf{e} to the variable \textsf{x}. In the
\emph{assume} operator, \assume{c}, if the boolean expression
\textsf{c} evaluates to \true, then the program continues, otherwise
it halts.  The set of operations is denoted by \typeop.
We then model a program by a \emph{transition system}. A transition
system is a quadruple $\quadruple{\locations}{\pcstart}{\trans}{O}$
where $\locations$ is the set of program points and
$\pcstart \in \locations$ is the
\emph{unique} initial program point. 
$\trans \subseteq \locations \times \locations \times \typeop$ is
the transition relation that relates a state to its (possible)
successors executing operations. This transition relation models the
operations that are executed when control flows from one program
point to another. We shall use
$\transition{\loc}{\loc'}{\textsf{op}}$ to denote a transition
relation from $\loc \in \locations$ to $\loc' \in \locations$ executing
the operation $\textsf{op} \in \typeop$. Finally, $O \subseteq
\locations$ is the set of terminal program points.

\vspace{2mm}
\noindent \textbf{Symbolic Execution (SE).} 
%
%
%
%
%
%
%
A \emph{symbolic} \emph{state} $\symstate$ is usually defined as a
triple $\triple{\state}{\store}{\pathcond}$. The symbol $\loc \in
\locations$ corresponds to the current program point. 
The symbolic store $\store$ is a function from program variables to
terms over input symbolic variables.  The \emph{evaluation}
$\eval{c}{\store}$ of a constraint expression $c$ in a store $\store$
is defined as: $\eval{x}{\store} = s(x)$ (if $x$ is a variable),
$\eval{n}{\store} = n$ (if $n $ is an integer), $\eval{e~{\sf
op}~e'}{\store} = \eval{e}{\store}~{\sf op}~{\eval{e'}{\store}}$
(where $e,e'$ are expressions and \textsf{op} is a relational or
arithmetic operator).  $\pathcond$ is called \emph{path condition} and
it is a first-order formula over the symbolic inputs and it
accumulates constraints which the inputs must satisfy in order for an
execution to follow the particular corresponding path. The set of
first-order formulas and symbolic states are denoted by \typefo\
and \typesymbstate, respectively.

For all purposes of this paper, we do not consider {\em arbitrary}
symbolic states, but only those generated during our symbolic
execution. For technical reasons, we require a symbolic state to be
aware of how it is reached in the symbolic execution tree.  Hence, we
abuse notation to (re)define a symbolic state as follows.

\begin{definition}[Symbolic State]
A symbolic state $\symstate$ is a quadruple
$\quadruple{\pc}{\store}{\pathcond}{\sympath}$, where $\pc$, $\store$,
$\pathcond$ are as before while the additional parameter $\sympath$ is
a sequence of program transitions that were taken during Symbolic Execution
in order to reach $\symstate$. $\Box$
\end{definition}
%
%

\begin{definition}[Transition Step]
Given a transition system $\quadruple{\locations}{\pcstart}{\trans}{O}$ and a
state $\symstate \equiv \quadruple{\pc}{\store}{\pathcond}{\sympath} \in \typesymbstate$,
the symbolic execution of $\transition{\state}{\state'}{\textsf{op}}$
returns another symbolic state $\symstate'$ defined as: \\

$\SYMSTEP(\symstate, \transition{\pc}{\pc'}{\sf op}) \equiv \symstate' \define$ \\ 
\vspace{-4mm}
\begin{eqnarray} 
\hspace{-1mm}
\left\{
      \begin{array}{ll}
        \hspace{-2mm}\quadruple{\pc'}{\store}{\pathcond \wedge \eval{c}{\store}}{\sympath'}       &  \hspace{-1mm}
       \textup{if \textsf{op} $\equiv$ \assume{c} and $\pathcond \wedge \eval{c}{\store}$} 
       \\ &  \hspace{-1mm} \textup{is satisfiable} \\
        \hspace{-2mm}\quadruple{\pc'}{\store[x \mapsto
                    \eval{e}{\store}]}{\pathcond}{\sympath'}& 
        \hspace{-1mm} \textup{if \textsf{op} $\equiv$ \assign{x}{e}} \\
      \end{array} \right.
  \hspace{-8mm}\label{eq:symbexecstep}
  \end{eqnarray}

 \noindent where $\sympath' \define\ \sympath~\cdot~\transition{\pc}{\pc'}{\sf op}$.
 We call $\symstate'$ a \emph{successor} of $\symstate$. $\Box$
\end{definition}

 Note that Eq.~(\ref{eq:symbexecstep}) queries a \emph{theorem prover}
 for satisfiability checking on the path condition. In practice, we
 assume the theorem prover is sound but not necessarily complete. That
 is, the theorem prover must say a formula is unsatisfiable only if it
 is indeed so.  Given a symbolic state
 $\symstate \equiv \quadruple{\pc}{\store}{\pathcond}{\sympath}$ we
 define $\mapstatetoformula{\symstate}:\typesymbstate~\next~\typefo$
 as the
  formula $
  (\bigwedge_{v~\in~\typevar}~\eval{v}{\store}) \wedge \pathcond$
  where $\typevar$ is the set of program variables.
%
  Such projection step is performed by eliminating existentially all
  auxiliary variables that are not in $\typevar$. As a convention, we
  use $\unknown$ in a tuple to denote a value that we are not
  interested in.

  A \emph{symbolic path} $\symstate_{0} \cdot \symstate_{1}
  \cdot ... \cdot \symstate_{n}$ is a sequence of symbolic states such
  that $\forall i \bullet 1 \leq i \leq n$ the state $\symstate_{i}$
  is a \emph{successor} of $\symstate_{i-1}$.
  %
  %
  A path $\symstate_{0} \cdot \symstate_{1} \cdot ... \cdot
  \symstate_{n}$ is \emph{feasible} if $\symstate_{n} \equiv
  \quadruple{\pc}{\store}{\pathcond}{\unknown}$ such that
  $\eval{\pathcond}{\store}$ is satisfiable. If $\pc \in O$ and
  $\symstate_{n}$ is feasible then $\symstate_{n}$ is called
  \emph{terminal} state, denoted \TERMINAL($\symstate_n$). Otherwise, if
  $\eval{\pathcond}{\store}$ is
  unsatisfiable the path is called \emph{infeasible} and
  $\symstate_{n}$ is called an \emph{infeasible} state, denoted
  \INFEASIBLE($\symstate_n$). 
  \ignore{A state
  $\symstate \equiv \triple{\pc}{\unknown}{\unknown}$ is
  called \emph{subsumed} if there exists another state
  $\symstate' \equiv \triple{\pc}{\unknown}{\unknown}$ such that
  $\mapstatetoformula{\symstate} \models \mapstatetoformula{\symstate'}$.}
  If there exists a feasible path $\symstate_{0} \cdot
  \symstate_{1} \cdot ... \cdot \symstate_{n}$ then we say
  $\symstate_{k}$ ($0 \leq k \leq n$) is \emph{reachable} from
  $\symstate_{0}$ in \emph{k steps}. We say $\symstate'$ is reachable
  from $\symstate$ if it is reachable from $\symstate$ in some number
  of steps.
  A \emph{symbolic execution tree} contains all the execution paths
  explored during the symbolic execution of a transition system by
  triggering Eq.~(\ref{eq:symbexecstep}). The nodes represent symbolic
  states and the arcs represent transitions between states.

  \begin{assumption}
  Given a terminal state
  $\symstate \equiv \quadruple{\pc}{\store}{\pathcond}{\sympath}$, we
  {\em assume} the existence of a function $\theta$ which extracts
  from $\sympath$ an ``exact'' analysis $\theta(\sympath)$. $\Box$
  \end{assumption}

  This exactness is a theoretical concept that helps us quantify the
  precision of our incremental analysis against a fully path- and 
  context-sensitive algorithm. We note that such algorithms do exist for
  loop-free programs, but they often do not work in the setting of
  realistic memory/timing budget.

  %
  \ignore{ We say a symbolic execution tree is \emph{complete} if it is finite
  and all its leaves are either terminal, infeasible or subsumed.
  }

\vspace{2mm}
\noindent{\bf Abstract Interpretation (AI).}
An invocation of AI at a symbolic state
$\symstate$ constructs an AI node at $\symstate$,
using an abstract domain \absdomain, which is assumed to 
be a lattice. This step starts by making use of the abstraction function \absfunc\
to map the current symbolic state $\symstate$ to
an abstract value, i.e., computing \absfunc(\symstate), then 
performing a standard AI computation over  $\absdomain$ and the
input CFG.

\begin{assumption}
We expect an invocation of AI would necessarily return an {\em upper
bound} analysis $\UB$, i.e., a safe over-approximation, of the set of
paths through $\symstate$.  In addition, we also assume that it
produces {\em witness paths} denoted as $\witness{}$ -- a set of paths
from which the upper bound analysis $~\UB$ is derived.
$\Box$
\end{assumption}

For WCET analysis, $\witness{}$ contains only one path. But for
general analyses, $\witness{}$ often is just a small subset of all the
paths going through $\symstate$, since not all paths contribute to the
returned analysis $\UB$.

\begin{assumption}
We assume the analysis values also form a lattice structure ${\cal
R}$, or $\quintuple{{\cal
S},\sqsubseteq}{\bot}{\sqcup}{\sqcap}{\top}$, where ${\cal S}$ is the
set of analysis values, $\sqsubseteq$ is the partial order
relationship, $\sqcup$ and $\sqcap$ are the least upper bound and
greatest lower bound operators, and $\bot$ and $\top$ are the bottom
and top elements of the lattice respectively. We assume that $\sqcup$
is precise so that exact analyses over different paths can be combined
precisely, yielding an exact analysis over the collection of paths.
$\Box$
\end{assumption}

Note that unlike the abstract domain $\absdomain$ to run AI, ${\cal
R}$ does not concern (partially) the (in-)feasibility of the program
paths.  While $\absdomain$ is typically designed to be coarse enough
so that an invocation of AI is fast, ${\cal R}$ is designed with
precision in mind. In practice, this does not hamper the overall
scalability of our algorithm, since operations on the analysis values
defined by ${\cal R}$ are performed only a small number of times,
bounded by the number of nodes in the HSET. (In other words, this 
assumption does not add any extra complexity to our algorithm.)


To simplify the presentation of the hybrid symbolic execution tree
(HSET), we also assume that an invocation of AI also returns a {\em
lower bound} analysis $\LB$.  In practice, one can always make use of
a trivial lower bound, e.g., $\bot$.  Let $\mathcal{E}$ be the desired
{\em exact} analysis of the set of paths, then
$\LB \sqsubseteq \mathcal{E} \sqsubseteq \UB$ (we do not explicitly
compute $\mathcal{E}$ but infer it when $\LB$ and $\UB$ coincide).

It now should be clear that under our notion of exactness and the
above assumptions, the analysis for each terminal symbolic state
$\symstate \equiv \quadruple{\pc}{\store}{\pathcond}{\sympath}$ is
exact. Its lower bound and upper bound coincide at $\theta(\sympath)$.

We comment here that computing the witness paths is straightforward
but tedious, so we do not detail it here. Typically an AI algorithm
operates over a CFG.  During its execution, it can ``mark'' certain
edges in the CFG that are sufficient to produce the analysis of the
target AI node where the analysis is invoked.  The witness paths can
then be obtained by traversing marked edges from the target AI node to
a terminal node (of the CFG). One can also follow \cite{cerny13popl}
to get not only the upper bound analysis, but also the {\em extremal} path(s)
from an abstract computation (within an iteration).
In summary, it is reasonable to assume
the existence of a procedure \algoai\ which when invoked with a
symbolic state $\symstate$ returns a triple
$\tuple{\LB}{\UB}{\witness{}}$.

\vspace{2mm}
\noindent
\textbf{Interpolation}. Given a pair of first order logic formulas $A$
and $B$ such that $A \wedge B$ is $false$, an
\emph{interpolant}~\cite{craig55interpolant} is
another formula $\Intpsymbol$ such that (a) $A \models \Intpsymbol$, (b)
$\Intpsymbol \wedge B$ is  $\false$, and (c) $\Intpsymbol$ is formed using
common variables of $A$ and $B$.  An interpolant removes irrelevant information
in $A$ that is not needed to maintain the unsatisfiability of $A \wedge B$.

Interpolation has been prominently used to reduce state space blowup
in both program verification~\cite{mcmillan10cav,jaffar09intp}
and program analysis~\cite{chu11emsoft,jaffar12sas}.  Here we
will use it for a similar purpose -- to merge, or {\em subsume},
symbolic states and avoid redundant exploration. During symbolic
execution, our algorithm will annotate certain states with an
interpolant, which can be used to prune other symbolic trees.
follows.

\begin{definition}[Subsumption]
Given a current symbolic state $\symstate \equiv
\quadruple{\pc}{\store}{\unknown}{\unknown}$
and an already explored symbolic state at the same program point $\symstate'
\equiv
\quadruple{\pc}{\unknown}{\unknown}{\unknown}$ annotated with the interpolant
\Intpsymbol,
we say $\symstate'$ \emph{subsumes} $\symstate$, denoted as
$\SUBSUMES(\symstate', \symstate)$ if (a) $\eval{\symstate}{\store} \models
\Intpsymbol$ and (b) $\absfunc(\symstate) \sqsubseteq \absfunc(\symstate')$.
%
\end{definition}

The first condition ensures that the symbolic paths through
$\symstate$ are a {\em subset} of the symbolic paths through
$\symstate'$, and the second condition ensures that the HSET at
$\symstate'$ has already been explored with a more general context
$\absfunc(\symstate')$. Therefore, by exploring $\symstate$ one cannot
obtain a more precise analysis than that has been already obtained by
exploring $\symstate'$, and hence $\symstate$ can be subsumed.

We note that subsumption is a special form of reuse that has been
briefly discussed in the early Sections. While \emph{reuse} (with
interpolation) has been exploited for different analysis
problems~\cite{chu11emsoft,jaffar12sas}, formulating this concept for
a \emph{general} analysis framework is rather involved.  For
simplicity, we thus omit the detail.

To conclude this Section, we comment that efficient interpolation
algorithms do exist for quantifier-free fragments of theories such as
linear real/integer arithmetic, uninterpreted functions, pointers and
arrays, and bitvectors (e.g., see~\cite{Cimatti-Tacas08} for details)
where interpolants can be extracted from the refutation proof in
linear time on the size of the proof.

\section{Algorithm}
\label{sec:algorithm}
\newcounter{proglineno}
\newcommand{\putno}{\refstepcounter{proglineno}\arabic{proglineno}:}

\begin{figure*}
\hspace{-0.5cm}
\begin{tabular}{ll}

\begin{tabular}{l}
\begin{minipage}{0.5\textwidth}
\begin{tabbing}
xxx \= x \= x \= x \= x \= x\=\kill
\setcounter{proglineno}{0}

\algoincr~($P$) \\

\putno\label{incr-init} \> $\symstate\ \Assign\ \quadruple{\pcstart}{\emptyset}
            {\true}{\unknown}$ \\

\putno\label{incr-ai} \> $\triple{\LB}{\UB}{\witness{}}$ \Assign\
            \algoai~(\symstate) \\

\putno\label{incr-do} \> \Do \\

\putno\label{incr-r} \>\> $R\ \Assign\ \{\symstate ~|~ \nexists\
            \symstate'$ s.t. $\DOMINATES(\symstate', \symstate)\}$ \\

\putno\label{incr-heu} \>\> $\symstate\ \Assign\ \algorefineheu\ (R)$ \\

\putno\label{incr-let} \>\> \Let\ $\quadruple
            {\LB}{\UB}{\witness{}}{\Intpsymbol}$ be the annotation at $\symstate$\\

\putno\label{incr-select-witness} \>\> {\bf select} a witness path $\awitness{\symstate}$ from $\witness{}$ \\

\putno\label{incr-ref} \>\> 
	{\sf spine\_done} \Assign\ \false;
            \algorefine(\symstate, $\awitness{\symstate}$) \\

\putno\label{incr-prop} \>\> \algopropagate(\symstate) \\

\putno\label{incr-until} \> \Until\ \algoboundsheu \\

\end{tabbing}
\end{minipage}

\\

\begin{minipage}{0.5\textwidth}
\begin{tabbing}
xxx \= x \= x \= x \= x \= x\=\kill

\algopropagate~($\symstate' \equiv
\quadruple{\pc}{\unknown}{\unknown}{\unknown}$) \\

\putno\label{prop-root} \> \If\ $\pc\ \equiv\ \pcstart$ \Then\ \Return \\

\putno\label{prop-init} \> \Let\ $\symstate$ be the predecessor of  $\symstate'$ \\

\putno\label{prop-init1} \> $\quadruple{\LB}{\UB}{\witness{}}{\Intpsymbol}
            \Assign\ \quadruple{\bot}{\bot}{\emptyset}{\true}$ \\

\putno\label{prop-for} \> \Foreach\ successor $\symstate''$ of $\symstate$ wrt. the transition $\transition{\pc}{\pc''}{\sf op}$ \\

\putno\label{prop-let} \>\> \Let\ $\quadruple{\LB''}{\UB''}{\witness{}''}{\Intpsymbol''}$ be the annotation at $\symstate''$ \\

            
\putno\label{prop-lub}  \> \> $\triple{\LB}{\UB}{\witness{}}$
            \Assign~          \algocombine($\triple{\LB}{\UB}{\witness{}}$, $\triple{\LB''}{\UB''}{\witness{}''}$)  \\

\putno\label{prop-wlp} \>\> $\Intpsymbol\ \Assign\ \Intpsymbol \wedge 
            \wpc~(\Intpsymbol'', {\sf op})$ \\

\putno\label{prop-endfor} \> \Endfor \\

\putno\label{prop-replace} \> replace $\symstate$'s annotation with
            $\quadruple{\LB}{\UB}{\witness{}}{\Intpsymbol}$ \\

\putno\label{prop-prop} \> \algopropagate($\symstate$) \\

\end{tabbing}
\end{minipage}

\\

\begin{minipage}{0.5\textwidth}
\begin{tabbing}
xxx \= x \= x \= x \= x \= x\=\kill

\algocombine~($\triple{\LB_1}{\UB_1}{\witness{1}}$, $\triple{\LB_2}{\UB_2}{\witness{2}}$) \\

\putno \> $\LB$ \Assign\ $\LB_1$ $\sqcup$ $\LB_2$ \\
\putno \> \If ~$\UB_1  ~\sqsubseteq ~\UB_2$ ~\Then ~$\witness{} ~\Assign ~\witness{2}$ \\
\putno \> \Else ~\If ~$\UB_2  ~\sqsubseteq ~\UB_1$ ~\Then ~$\witness{} ~\Assign ~\witness{1}$ \\
\putno \> \Else ~$\witness{}$ ~\Assign ~$\witness{1} ~\cup ~\witness{2}$ \\
\putno \> $\UB$ ~\Assign~ $\UB_1$ ~$\sqcup$ ~$\UB_2$ \\
\putno \> \Return ~$\triple{\LB}{\UB}{\witness{}}$ \\

\end{tabbing}
\end{minipage}

\end{tabular}

&

\hspace{-0.5cm}
\begin{minipage}{0.5\textwidth}
\begin{tabbing}
xxx \= x \= x \= x \= x \= x\=\kill

\algorefine~($\symstate \equiv \quadruple{\pc}{\unknown}{\unknown}
{\sympath}$, $\awitness{\symstate}$) \\

\putno\label{ref-inf} \> \If\ \INFEASIBLE~(\symstate) \Then\\

\putno\label{ref-inf1} \>\> $\quadruple{\LB}{\UB}{\witness{}}{\Intpsymbol}
            \Assign\ \quadruple{\bot}{\bot}{\emptyset}{\false}$; {\sf spine\_done} \Assign\ \true;  \\

\putno\label{ref-term} \> \Else\ \If\ \TERMINAL~(\symstate) \Then\ \\

\putno\label{ref-term1} \>\> $\quadruple{\LB}{\UB}{\witness{}}{\Intpsymbol}
            \Assign\ \quadruple{\theta(\sympath)}
            {\theta(\sympath)}{\emptyset}{\true}$; {\sf spine\_done} \Assign\ \true \\

\putno\label{ref-sub} \> $\Else\ \If\ (\exists\ \symstate' \equiv \quadruple{\pc}
            {\unknown}{\unknown}{\unknown}$ s.t. $\symstate'$ is annotated 
            $\quadruple{\LB'}{\UB'}{\witness{}'}{\Intpsymbol'}$  \\
~~~~~~~~~~~~~~~~~~~~~~~~~~~~~~~~~~~~~ and $\SUBSUMES~(\symstate', \symstate))$  \Then\ \\         
\putno\label{ref-sub1} \>\> $\quadruple{\LB}{\UB}{\witness{}}{\Intpsymbol}$ \Assign\ $\quadruple{\LB'}{\UB'}{\witness{}'}{\Intpsymbol'}$; {\sf spine\_done} \Assign\ \true\ \\

\putno\label{ref-spine0} \> \Else\ \If\ {\sf spine\_done} \Then\  \\

\putno\label{ref-spine1} \>\> $\triple{\LB}{\UB}{\witness{}}$ \Assign\ $\algoai
            (\symstate)$; ~$\Intpsymbol\ \Assign\ \true$ \\

\putno\label{ref-else} \> \Else \\

\putno\label{ref-init} \>\> $\quadruple{\LB}{\UB}{\witness{}}{\Intpsymbol}
            \Assign\ \quadruple{\bot}{\bot}{\emptyset}{\true}$ \\
\putno\label{ref-wit-start} \> \> {\bf select} a transition s.t. $\transition{\pc}{\pc'}{\sf op} \in\ \awitness{\symstate}$ \\
\> \> \>  \> \> \>   \mbox{// There is only one such transition} \\

\putno \>\> $\symstate'\ \Assign\ \SYMSTEP(\symstate,
            \transition{\pc}{\pc'}{\sf op})$ \\
\putno\label{ref-spine} \>\> $\algorefine~(\symstate', \awitness{\symstate})$ ~ \mbox{// Target refinement towards $\awitness{\symstate}$} \\

\putno \>\> \Let\ $\quadruple{\LB'}{\UB'}{\witness{}'}{\Intpsymbol'}$ be the annotation of $\symstate'$   \\

            
\putno            \> \> $\triple{\LB}{\UB}{\witness{}}$
            \Assign~          \algocombine($\triple{\LB}{\UB}{\witness{}}$, $\triple{\LB'}{\UB'}{\witness{}'}$) \\

\putno\label{ref-wit-end} \>\> $\Intpsymbol\ \Assign\ \Intpsymbol \wedge 
            \wpc~(\Intpsymbol', {\sf op})$ \\
            
\putno\label{ref-for} \>\> \Foreach\ transition s.t. $\transition{\pc}{\pc'}{\sf op} \in\ P$ and   $\transition{\pc}{\pc'}{\sf op} \not\in\ \awitness{\symstate}$ \\

\putno\label{ref-symstep} \>\>\> $\symstate'\ \Assign\ \SYMSTEP(\symstate,
            \transition{\pc}{\pc'}{\sf op})$ \\

\putno\label{ref-ref} \>\>\> $\algorefine~(\symstate', \awitness{\symstate})$ 
\ptab \mbox{// An AI node will be built}
\\
            
\putno\label{ref-let} \>\>\> \Let\ $\quadruple{\LB'}{\UB'}{\witness{}'}{\Intpsymbol'}$ be the annotation of $\symstate'$   \\

            
\putno\label{ref-lub}            \> \> \> $\triple{\LB}{\UB}{\witness{}}$
            \Assign~          \algocombine($\triple{\LB}{\UB}{\witness{}}$, $\triple{\LB'}{\UB'}{\witness{}'}$) \\

\putno\label{ref-wlp} \>\>\> $\Intpsymbol\ \Assign\ \Intpsymbol \wedge 
            \wpc~(\Intpsymbol', {\sf op})$ \\

\putno\label{ref-endfor} \>\> \Endfor \\

\putno\label{ref-endif} \> \Endif \\
\putno \> remove the annotation of $\symstate$ \\
\putno\label{ref-store} \> \If\ $\LB\ \equiv\ \UB$ \Then\ annotate $\symstate$
            with $\quadruple{\LB}{\UB}{\emptyset}{\Intpsymbol}$ \\

\putno\label{ref-store1} \> \Else\ annotate $\symstate$ with $\quadruple{\LB}
            {\UB} {\witness{}}{\false}$ \\

\putno\label{ref-end} \> \Endif \\

\end{tabbing}
\end{minipage}

\end{tabular}
\caption{Algorithm for Incrementally Precise Analysis}
\label{fig:algo}
\vspace{-4mm}
\end{figure*}

Our incremental analysis algorithm, whose pseudocode is shown in
Fig.~\ref{fig:algo}, can be expressed as one that starts with an abstract
interpretation (AI) node representing an abstract analysis of the whole program, and
gradually refines the HSET using symbolic execution (SE) until the desired
level of analysis precision is obtained. Since each node in the HSET
corresponds to a symbolic state $\symstate$, we will call it node $\symstate$ for short.
During SE, a forward traversal collects
path constraints and checks for path feasibility, and a backtracking phase {\em
annotates} each node $\symstate$ in the HSET with the following information:
$\quadruple{\LB}{\UB}{\witness{}}{\Intpsymbol}$, representing the lower bound
and upper bound analyses for the set of paths through $\symstate$, the 
witness paths for the upper bound analysis, and the interpolant at $\symstate$,
respectively.

With this annotation, we now define our all important {\em domination}
condition.

\begin{definition}[Domination]
  A node $\symstate$ annotated with $\quadruple{\LB}{\UB}{\witness{}}
  {\Intpsymbol}$ is {\em dominated} by a node $\symstate'$
  annotated with $\quadruple{\LB'}{\UB'}{\witness{}'}{\Intpsymbol'}$ if\ ~$\UB\
  \sqsubseteq\ \LB'$. We also say that $\symstate'$ {\em dominates} $\symstate$,
  denoted as \DOMINATES($\symstate'$, $\symstate$).
  $\Box$
\end{definition}

In other words, if a symbolic state produces an upper bound analysis that is
already {\em contained} (lattice-wise) in the lower bound analysis of another
state, it is considered dominated. Particularly, there is no use trying to
refine it to reduce its upper bound analysis. Note that a node can dominate
itself if its lower and upper bounds are the same (i.e., it has an {\em exact}
analysis). Obviously a node with an \emph{exact} analysis does not need
to be refined further.

The main procedure, \algoincr, accepts the program $P$ as a transition
system, which we assume is a global variable to all procedures.  In
line~\ref{incr-init}, the initial state is created with $\pcstart$ as
the program point, an empty store, the path condition \true, and the
empty sequence.  In line~\ref{incr-ai} the initial HSET containing a
single AI node is generated by calling
\algoai\ with the initial state. This would return a possible lower bound, an
upper bound and the witness paths $\witness{}$ for the upper bound.

Lines~\ref{incr-do}-\ref{incr-until} define the main refinement
loop. Our choice of AI node to refine, in conjunction with building
the spine targeting the witness paths, is what makes our algorithm
\emph{goal-directed}:

\vspace*{3mm}
\hspace*{3mm}\begin{minipage}{75mm}
choose an AI node which is (a) \emph{not dominated}, and (b) has a
maximal upper bound.
\end{minipage}
\vspace*{3mm}

\noindent
In the algorithm, first the set {\em non-dominated} AI nodes in the
current HSET is collected in $R$.  Choosing a node with \emph{maximal}
upper bound analysis, in the case of WCET, is easy because the
analysis values range over positive integers.  In other analyses, if
possible, a ``difference'' metric can be defined to even measure the
amount of (non) domination, and the AI node in $R$ with maximal
difference can be chosen.

\bigskip
\noindent
{\bf Remark: } Before continuing with the description of the
algorithm, let us comment on the properties of our choice of
refinement.  Let $N$ be an AI node which is (a) \emph{not dominated},
and (b) has a maximal upper bound among other non-dominated AI nodes.
Refining any other node, say $M$, will increase $M$'s lower bound of
decrease $M$'s upper bound. However, (1) $N$ will never become dominated,
and (2) the overall analysis will not be improved.  We note further
that while (2) relies on the assumption that the analysis values
constitute a lattice, (1) holds even when we relax that assumption,
allowing the analysis values to only be a semi-lattice.
(End of Remark.)

\bigskip
Once the AI node $\symstate$ is chosen for refinement,
the procedure \algorefine\ is called along with the witness paths for
its upper bound analysis $\witness{}$. When \algorefine\ returns it
would have annotated $\symstate$ with new, possibly tighter, upper and
lower bounds which are then propagated back to its ancestors by the
procedure \algopropagate. This process continues until the loop
terminates by means of a \algoboundsheu, which is user-defined.

A straightforward \algoboundsheu\ check is to check if there are no
non-dominated symbolic states. This forces the algorithm to terminate
only when an {\em exact} analysis is derived. However, a WCET analyzer
could be content if, say, the difference between upper and lower
bounds is less than 5\%, in which case the heuristic can check if the
root of the HSET (the initial state) is annotated with
$\quadruple{\LB}{\UB}{\unknown}{\unknown}$ s.t. $(\UB\ - \LB) / \UB\
\leq\ 0.05$.

\ignore{
A taint analyzer may only care about the flow of taint to a particular
subset of variables $S$, so the heuristic can check if $\forall
\quadruple{\unknown}{\UB}{\unknown}{\unknown}, S \nsubseteq
\UB$, or if $\exists \quadruple{\LB}{\unknown}{\unknown}{\unknown}$ s.t. 
$S \subseteq \LB$, in order to ensure no-flow or flow of taint to the variables in $S$, respectively.
}

\algorefine\ is our main refinement procedure that accepts the current 
node $\symstate$ and the set of witness paths $\witness{\symstate}$.
It is a recursive procedure that refines an AI node by symbolically
unfolding the paths in $\witness{\symstate}$, with the hope of either
confirming or refuting the current analysis of the node.  There are
four bases of this procedure:

\begin{itemize}

\item (Lines~\ref{ref-inf}-\ref{ref-inf1}) If \symstate\ is an infeasible state,
then it sets the lower and upper bounds to $\bot$, the set of witness paths for
the upper bound to $\emptyset$, and the interpolant $\Intpsymbol$ to \false\ to
denote the infeasibility.

\item (Lines~\ref{ref-term}-\ref{ref-term1}) If \symstate\ is a terminal state,
then an {\em exact analysis} for this symbolic path is achieved. Hence both the lower and upper
bounds are set to $\theta(\sympath)$ -- the analysis extracted from
this single path. 
The witness paths for this analysis
can be set to $\emptyset$ because we will never refine an exact analysis in
future.  Finally, the interpolant is set to \true. In addition, we set a
(global) variable {\sf spine\_done} to \true\ to signify that a spine (witness
path) has been exercised fully, and can begin constructing AI nodes along the
branches from this path later.

\item (Lines~\ref{ref-sub}-\ref{ref-sub1}) If \symstate\ is subsumed by another
state $\symstate'$, it simply sets {\sf spine\_done} to \true. Implicitly, the
lower and upper bounds, the witness paths and interpolant for $\symstate$ are
copied over from $\symstate'$.

\item (Lines~\ref{ref-spine0}-\ref{ref-spine1}) If {\sf spine\_done} is \true,
i.e., a spine has been explored already and we are exploring other branches from
it, then it constructs an AI node at $\symstate$ by calling \algoai. This would
return a lower bound, upper bound and the witness paths for the upper bound. The
interpolant is then set to \true, as there is no infeasibility to capture in the
constructed AI node.

\end{itemize}


\noindent
If the four bases fail, \algorefine\ proceeds to the successors of
\symstate~(lines~\ref{ref-else}-\ref{ref-endif}). It first initializes the lower
and upper bounds, the witness paths, and interpolant to $\bot$, $\emptyset$ and
\true\ respectively, which will be modified.
Then we target the refinement to either confirm or refute the given witness path $\awitness{\symstate}$
(lines~\ref{ref-wit-start}-\ref{ref-wit-end}).
This is done by following the witness, applying $\SYMSTEP$ on $\symstate$ to construct the  next
symbolic state $\symstate'$. Then \algorefine\ is called recursively. 
For each remaining transition, which is not part of the witness path, the algorithm proceeds similarly 
(lines~\ref{ref-for}-\ref{ref-endfor}). But note that, now a spine has been constructed,
indicated by {\sf spine\_done} being set to \true, a number of AI nodes
will be computed along the spine. We further comment that a typical AI algorithm, 
when invoked, will follow the input CFG and compute an analysis for each program point,
not just for the point of invocation.
Thus the number of AI invocations 
while seemingly overwhelming, can indeed be optimized by a simple \emph{caching}
mechanism. In our implementation of our practical applications in 
Section~\ref{sec:experiments}, this is never an issue.

We now detail on how the analysis answer and the interpolant are aggregated.
Upon returning from the recursive call, $\symstate'$ would have been annotated
with some lower and upper bounds, witness paths, and interpolant.  From this,
the same information for $\symstate$ is computed by {\em joining} it with the
existing information at $\symstate$ (line~\ref{ref-lub}) using the
straightforward \algocombine\ procedure. That is, the analysis
of the set of paths through $\symstate$ is computed as the (lattice) join of the
analysis of each individual path. The interpolant
deserves some special treatment due to its back propagation. From the
interpolant $\Intpsymbol'$ at $\symstate'$, the interpolant at
$\symstate$ is computed by conjoining the current interpolant $\Intpsymbol$ with
$\wpc(\Intpsymbol', {\sf op})$ --- the {\em weakest liberal
precondition}~\cite{dijkstrawp} of $\Intpsymbol'$ w.r.t. the transition
{\sf op}. $\wpc : \typefo \times \typeop~\next~\typefo$ ideally
returns the weakest formula on the current state such that the execution of {\sf
op} results in $\Intpsymbol'$. In practice we approximate the \wpc{} by
making a linear number of calls to a theorem prover, using techniques outlined
in~\cite{jaffar09intp}, which usually results in a formula stronger than
\wpc{}.

Finally, once either a base case or the recursive case is executed, \algorefine\
annotates (lines~\ref{ref-store}-\ref{ref-end}) the current state with the
information defined by one of the cases. An important check is made here: if the
lower and upper bounds are the same, then we have an {\em exact} analysis at
$\symstate$. Therefore, the witness paths can be set to $\emptyset$ since we
will never refine an exact analysis. But most importantly, if the check {\em
failed}, then the bounds do not coincide, and the analysis is imprecise. A
state with an imprecise analysis should {\em not} subsume any other state. Hence
we change the interpolant to \false\ before annotating $\symstate$ so that for
all states $\symstate''$, $\SUBSUMES(\symstate, \symstate'')$ would fail. A
subtle corollary of this is that the first three base cases assign the same
lower and upper bounds at $\symstate$, and the fourth base case (AI) 
usually assigns them different values. The recursive case is then dependent on the
the bounds of the successors of $\symstate$.

The final procedure \algopropagate\ simply propagates the annotation at a given
state $\symstate'$ to its ancestors upto the root of the entire tree at
$\pcstart$. In line~\ref{prop-init}, it obtains the parent state $\symstate$,
and in lines~\ref{prop-init1}-\ref{prop-endfor} it performs the backward
propagation from all successors of $\symstate$, in exactly the same way as
lines~\ref{ref-init},\ref{ref-let}-\ref{ref-wlp} of \algorefine. For
brevity, we provide its pseudocode but omit a detailed description.

The whole algorithm is guaranteed to terminate provided \algoai\ terminates (see
discussion on unbounded loops below). 
In case the algorithm is interrupted and
forced to terminate, the current lower bound and upper bound can be extracted
easily from the symbolic states and presented to the user, making
this an ``anytime algorithm''.

\ignore{
We remark that {\em bounded loops} cause no problem to our described
algorithm.  For example, in the domain of WCET analysis, most loops
are required to be statically bounded. A common treatment in this
analysis domain is that loops are \emph{statically} unrolled. In this
case, all the technical description so far is very much applicable, as
the program's SET is finite.
}


\ignore{
\Vijay{TODO: CLEANUP THIS}

\noindent
{\bf Backward Analyses} such as WCET or slicing can be easily implemented using
our algorithm with a few changes. Firstly, the initial analysis provided to
\algoincr\ would be defined at \pcend\ rather than \pcstart. \algoai\ would
start from $\pcend$ and apply the $\pre$ operation rather than $\post$
(line~\ref{ai-post}).
\algorefine\ would still execute paths in a forward manner in order to detect
infeasibility, but the analysis propagation (line~\ref{ref-post}) would again
apply the $\pre$ operation on $\answer{\symstate'}$ to get $\answer{\symstate}$.

The most important change is the notion of {\em reuse} of backward analysis
answers. In forward analyses, when a state $\symstate$ is subsumed by
$\symstate'$, we implicitly noted that by exploring the tree rooted at
$\symstate$ one would never obtain a more precise analysis than that already
obtained because the analysis depended only on
$\answer{\symstate}$. In backward analyses, the analysis of the tree rooted at
$\symstate$ depends not on $\answer{\symstate}$ but on the {\em feasible paths}
in the tree.

\Vijay{END OF TODO}
}

\ignore{ 246

{\em Bounded Loops} cause no problem to our described algorithm.  For
example, in the domain of WCET analysis, most loops are required to be
statically bounded. A common treatment in this analysis domain is that
loops are \emph{statically} unrolled. In this case, all the technical
description so far is very much applicable, as the program's SET is
finite.

\vspace{2mm}
\noindent
{\em Unbounded Loops} pose a technical problem as they make the SET
infinite, thereby making \algorefine\ non-terminating. As we only
work with finite SE trees, the only possibility to get termination is by using 
an abstraction such as a loop invariant. We
employ invariant generation techniques outlined
in~\cite{jaffar09intp,jaffar12sas}. Particularly we assume that program points
are labeled with invariants inferred from an external invariant generator,
typically using abstract domains such as octagons or polyhedra~\cite{Seo03}, 
and a function \getassrt\ which,
given a program point $\pc$ and symbolic store $\store$, returns an assertion in
the form of a FOL formula, renamed using $\store$, that holds at $\pc$. Note
that when $\pc$ is a loop header, \getassrt\ will return a loop invariant. Then,
we add a new case to $\SYMSTEP$: \\

\noindent
$\SYMSTEP(\symstate, \transition{\pc}{\pc'}{\sf op})$
$\define\
\InvariantFunc(\tuple{\pc'}{\store}{\pathcond}, \pc \rightarrow \pc_n)~~~$ if
$\state$ is the header for a loop from $\pc$ to $\pc_n$ \\

\noindent
where $\InvariantFunc(\quadruple{\state}{\store}{\pathcond}{\sympath}, \state \rightarrow
\state_{n}) \define$

\vspace{-5mm}
  \begin{eqnarray*} 
     \left\{
      \begin{array}{l}
        \textup{\Let}~ \store'~\textup{\Assign}~\havoc(\store,\modifies(\state \rightarrow \state_{n})) \\
        ~~\hspace{3mm} \pathcondbar~\textup{\Assign}~ \getassrt(\state,\store') \wedge \pathcond \\
        \textup{\In}~\quadruple{\state}{\store'}{\pathcondbar}{\sympath}
        ~~\hspace{3mm}
      \end{array}  \right.
  \end{eqnarray*}

\noindent
  $\havoc(\store, Vars) \define \forall v \in Vars \bullet \store[ v
  \mapsto z ]$
  where $z$ is a fresh variable (implicitly $\exists$-quantified).

\noindent
  $\modifies(\state \rightarrow \ldots \rightarrow \state_{n})$
  takes a sequence of transitions and returns the set of variables
  that may be modified during its symbolic execution.

Intuitively, \InvariantFunc\ clears the symbolic store of all
variables modified in the loop (using the \havoc\ function) and then
strengthens the path condition $\Pi$ of the symbolic state with the
externally obtained invariants. These invariants, once and for all,
define the structure of the SET that will be explored iteratively by
our algorithm. Note that we do not attempt to refine these invariants
-- they are provided to us by an ``oracle''.

The notion of a spine is also affected by unbounded loops: a spine is
no longer just a straight symbolic path from the root to a terminal
point, but one that also contains looping points (cycles) at the loop
headers. We can still extract a lower bound analysis from this spine
by performing the standard fixpoint computation on the abstract domain
$\absdomain$, but undecidability restricts us from guaranteeing that
this analysis is {\em exact}. Therefore, dominations performed using
this lower bound may result in an imprecision of the overall analysis,
but soundness is still guaranteed. The imprecision is affected by the
quality (i.e., logical strength) of the loop invariants, but this
topic is outside the scope of this paper.

Finally, \algorefine\ is modified so that if it detects a cycle at
$\pc$, it simply stops unfolding and returns, as $\pc$ has already
been explored with the loop invariant, thus guaranteeing its
termination.

} 


\section{Experimental Evaluation}
\label{sec:experiments}
We implemented the incremental analysis algorithm in Fig.~\ref{fig:algo} on the
\tracer\ framework for symbolic execution, using the same
interpolation method and theory solver presented in~\cite{chu11emsoft}.  We
instantiated our algorithm for a backward WCET analysis.
%
%
The analysis values form the lattice 
$\mathcal{R}_1 \equiv \quintuple{\mathbb{N},\leq}{0}{\sqcup}{\sqcap}{\infty}$, with $\mathbb{N}$
is the set of non-negative integers, and $A_{1} \sqcup\ A_{2} \define\
max(A_{1}, A_{2})$. 
The abstract domain $\absdomain$ used for our AI component is 
the domain of intervals, 
which is well-known for its efficiency.

We implemented the heuristics in Fig.~\ref{fig:algo} as follows. \algorefineheu\
is quite straightforward as the lattice $\quintuple{\mathbb{N},\leq}{0}{\sqcup}{\sqcap}{\infty}$ 
imposes a total order on its elements. Hence we simply pick for refinement the AI node that
produced the \emph{maximum}\footnote{A maximum always exists.} upper bound WCET, with ties being resolved
non-deterministically. \algoboundsheu\ implements the following
check: 

\begin{center}
$\forall~\symstate ~ \exists~\symstate' ~s.t. ~\DOMINATES(\symstate', \symstate)$, 
\end{center}

\noindent
that is, every symbolic state is dominated by another state, possibly
by itself if it produces an exact analysis. This makes the algorithm terminate
only when the final WCET is exact.
\ignore{
We emulated the low-level cache using a coarse model of an ``instruction cache''.
The cache size is 32 instructions, and the replacement policy on a cache miss is
to simply erase the cache and fetch the next 32 instructions in the current
basic block. It takes 1 unit of time to execute a program statement, and the
cache miss penalty is 128 units of time.
}

We used a simple model of an ``instruction cache''.
It is a direct-mapped cache of size 4KB. Each cache set can hold 32 instructions.
It takes 1 unit of time to execute a program statement, and the
cache miss penalty is 128 units of time.

\ignore{ 753

\subsection{Taint Analysis}
For forward taint analysis, the analysis values form the lattice 
${\cal R}_2 \equiv \quintuple{\mathbb{P},\subseteq}{\emptyset}{\cup}{\cap}{V}$, with $\mathbb{P}$ 
is the powerset of $V$, the set of all program variables, and standard operators on
sets. The abstract domain $\absdomain$ is defined as
${\cal I} \times {\cal R}_2$, where ${\cal I}$, as before, is the domain of
intervals. In the literature, there are various definitions of how taint information is to be propagated, 
and we follow the one in~\cite{schwartz10sp}, which considers both both {\em explicit} and
{\em implicit} tainting.  Explicit taint occurs when there is a direct data-flow
from a tainted variable, say $t$, to another variable $x$ (e.g., through an
assignment {\tt x=t+r}). Implicit taint occurs when there is an {\em equality}
check on a tainted variable, such as {\tt if(t==x)}, which intuitively makes the
other argument $x$ also tainted, as one can observe its value to find the value
of $t$.


%
%

Typically, the source of taint affects the propagated taint information.
But for lack of any information regarding the source, we picked all variables
whose values were obtained from outside the program, including {\tt extern}
variables, user input, environment variables, etc., and tainted some random
subset of them. This models the usual scenario where some parts of the
environment are ``sensitive'' (such as a user's password input) and others are
not (such as the system time).


For \algorefineheu\,
unlike WCET, the lattice ${\cal R}_2$ does not impose a total order
since two variable sets may be incomparable.  Therefore we defined a notion of
``difference'' in domination: first, we compute the lattice join
($\sqcup$) of all lower bound taint sets in $L_B$.  Then, for each
$\symstate$ that roots an AI node annotated with
$\quadruple{\LB}{\UB}{\witness{}}{\Intpsymbol}$, we compute the difference set
$\delta_{\symstate} \equiv\ \UB\ \setminus\ L_B$. We then pick the AI node at
$\symstate$ that produces the maximal value of $| \delta_{\symstate} |$.  That
is, we pick the AI node that causes a maximum difference in the {\em
cardinality} of tainted variable sets compared to the collective lower bound.
Ties, when two
sets produce the same difference in cardinality, are resolved
non-deterministically.  For instance, if two AI nodes, at $\symstate$ and
$\symstate'$, produced the taint sets $\{a,b,c\}$ and $\{a,c,d\}$, and the
lower bound $L_B$ is $\{a,b\}$, then we pick $\symstate'$ to refine, as
it produces the difference $\{c,d\}$ as opposed to just $\{c\}$ from
$\symstate$.  Finally, \algoboundsheu\ implements the same check as for WCET
analysis, which ensures that the taint analysis is exact when the algorithm
terminates.

} 

\begin{table*}
\begin{center}
\begin{tabular}{|l|r||r||r|r|r||r|r|r|r||r||}
\hline
{\sf Benchmark} & {\sf LOC} & \multicolumn{1}{|c||}{\sf AI-based} & \multicolumn{3}{|c||}{\sf
Full SE~\cite{chu11emsoft}} & \multicolumn{4}{|c||}{\sf Incremental} & \multicolumn{1}{|c||}{\sf \% Imp} \\
\cline{3-10}

& & {\sf WCET} & {\sf WCET} & {\sf Time} & {\sf Mem} & ${\sf WCET}_\UB$ & ${\sf
WCET}_\LB$ &
{\sf Time} & {\sf Mem} &
 \\
\hline

{\sf cdaudio} & 1288 & 10663 & 9370 & 28 s & 212 MB & 9370 & 9370 & 14 s & 56 MB & 13.8\% \\

{\sf diskperf} & 1255 & 33598 & $\infty$ & $\infty$ & 2 GB & 29723 & 29723 & 231 s & 400 MB & 13.0\% \\

{\sf floppy} & 1524 & 16627 & 13784 & 19 s & 136 MB & 13784 & 13784 & 15 s & 44 MB & 20.1\% \\

{\sf ssh} & 2213 & 12394 & 6075 & 17 s & 39 MB & 6075 & 6075 & 17 s & 51 MB & 104\% \\

{\sf nsichneu} & 2540 & 206788 &  $\infty$ & $\infty$ & 522 MB & 52430 & 52430 & 156 s & 133 MB & 294\% \\

{\sf tcas} & 235 & 29305 & $\infty$ & $\infty$ & 1.4 GB & 28788 & 23887 & $\infty$ & 432 MB & 2\% \\

{\sf statemate} & 1187 & 31281 & $\infty$ & 285 s & $\infty$ & 31151 & 18623 & $\infty$ & 767 MB & 0.5\% \\

\end{tabular}
\end{center}
\vspace{-3mm}
\caption{WCET Analysis results for AI based, SE based, and our incremental
algorithm. An $\infty$ represents a timeout or out-of-memory.}
\label{tab:wcet}

\end{table*}

\ignore{

\begin{table*}
\hspace{1cm}
\begin{tabular}{|l|r||r||r|r|r||r|r|r|r||r||}

\hline
{\sf Benchmark} & {\sf \# V} & \multicolumn{1}{|c||}{\sf AI-based} & \multicolumn{3}{|c||}{\sf
SE~\cite{jaffar12sas}} & \multicolumn{4}{|c||}{\sf Incremental} & \multicolumn{1}{|c||}{\sf \% Imp} \\
\cline{3-10}

& & {\sf \# TV} & {\sf \# TV} & {\sf Time} & {\sf Mem} & ${\sf \#
TV}_\UB$ & ${\sf \# TV}_\LB$ &
{\sf Time} & {\sf Mem} &
 \\
\hline

{\sf cdaudio} & 330 & 50 & $\infty$ & $\infty$ & 574 MB & 45 & 45 & 17 s & 227 MB & 99\\

{\sf diskperf} & 185 & 36 & $\infty$ & $\infty$ & 425 MB & 31 & 31 & 7 s & 101 MB & 99\\

{\sf floppy} & 330 & 27 & $\infty$ & $\infty$ & 581 MB & 20 & 20 & 12 s & 229 MB & 99\\

{\sf ssh} & 63 & 57 & 53 & 1 s & 8 MB & 53 & 53 & 1 s & 8 MB & 99\\

{\sf nsichneu} & 22 & 16 & 16 & 4 s & 24 MB & 16 & 16 & 2 s & 11 MB & 99\\

{\sf tcas} & 41 & 15 & 15 & 4 s & 55 MB & 15 & 15 & 1 s & 12 MB & 99\\

{\sf statemate} & 119 & 67 & $\infty$ & $\infty$ & 770 MB & 63 & 63 & 7 s & 79 MB & 99\\

\end{tabular}
\caption{Taint Analysis results. {\sf \# TV} measures
the number of tainted variables. An $\infty$ represents a timeout or out-of-memory.}
\label{tab:taint}
\end{table*}

}

We used as benchmarks sequential C programs from a varied pool -- three device
drivers {\sf cdaudio}, {\sf diskperf}, {\sf floppy} from the {\sf
ntdrivers-simplified} category and SSH Client protocol from the {\sf
ssh-simplified} category of SV-COMP~2014~\cite{SVCOMP14}, an air traffic
collision avoidance system {\sf tcas}, and two programs from the M\"{a}lardalen
WCET benchmark~\cite{malardalenbenchmark} {\sf statemate} and {\sf nsichneu}. We
removed the safety properties from the SV-COMP benchmarks as we are not
concerned with their verification. All experiments are carried out
on an Intel 2.3 Ghz machine with 2GB memory, with a timeout of 5 minutes,
considering our nominal benchmark size. 
\ignore{
The lower timeout for taint analysis is because it is relatively cheaper to
perform than WCET, as will be evident from the timings in Table~\ref{tab:taint}.
This is because the abstract domain of taint analysis uses a {\em finite}
lattice bounded by the number of program variables, and hence convergence is
achieved easier when compared to WCET which uses an infinite lattice of
integers.
}

We compared our incremental algorithm with two
adversaries: abstract interpretation (AI) on one hand, and a state-of-the-art SE
based algorithm \cite{chu11emsoft} on the other. 
\ignore{
In particular, ee chose the algorithm
presented in~. For taint analysis, we modified the
algorithm in~\cite{jaffar12sas} to propagate forward taint information instead
of slice information. These algorithms are highly path-sensitive, designed to
produce {\em exact} analysis, and employ aggressive pruning techniques such as
interpolation and reuse to achieve scalability.
}
We present the following statistics, in Table~\ref{tab:wcet}, for each benchmark: (a)
the final analysis produced by the AI-based, SE-based, and our incremental
algorithm with upper ($\UB$) and lower ($\LB$) bounds (b) the time taken, (c)
the total memory usage as given by the underlying \tracer\ system,
and finally (d) collation of the previous columns into a 
\emph{imprecision improvement} {\sc \% Imp}.
This is defined as the percentage of $(A - I)/I$
where $A$ and $I$ are the AI-based and incremental analyses respectively.
%
%
\noindent
We do not show the time and memory for the AI based algorithm as they
are quite negligible compared to those of the other two
algorithms. For instance, it always terminates in less than 1 second.


The AI based algorithm produces an analysis quickly for all
programs as mentioned above, but it is in fact not precise. As we will see,
there is at least a 10\% precision improvement in most benchmarks, and an alarming 300\% in {\sf
nsichneu}, a well-known program in the WCET community that is particularly known
to be hard to analyze. So the only hope to
produce an exact analysis is if the SE based algorithm terminates.  However this
fails to terminate by either timing out or running out of memory for
four out of our seven benchmarks, leaving behind no useful analysis information.

On the other hand, our incremental algorithm is able to provide useful information.
In the first five benchmarks where it terminated within the budget,
it of course produced  an {\em exact} analysis, but in most cases,
it used much less than the allocated budget.
\ignore{
It either terminates well before the SE algorithm, as in all but the last two
programs, thus producing an {\em exact} analysis using much less budget (time
and memory), or it produces a more precise {\em range} for the analysis using
}
For the remaining two benchmarks, our algorithm produced a more precise {\em range}
for the analysis via tighter upper and lower bounds.
For instance, in {\sf nsichneu}, AI produced the imprecise WCET 206788 and the SE-based algorithm
ran out of budget.   However our algorithm was able to produce the exact WCET of 52430 in
less than half the budget.
Furthermore, it used only a quarter of the memory as did SE.
This seems to be a common trend across all our benchmarks
(with the exception of {\sf ssh}).

We do admit here that the improvement WCET our algorithm had produced
was small in the two benchmarks where termination was abruptly
enforced.  (But, importantly, it was not zero.)  Given the good performance of our
algorithm on all the other benchmarks where our algorithm did
terminate, we will speculate here that these two (nonterminating)
benchmarks may be unfortunate outliers.

\begin{figure}
\includegraphics[scale=0.7]{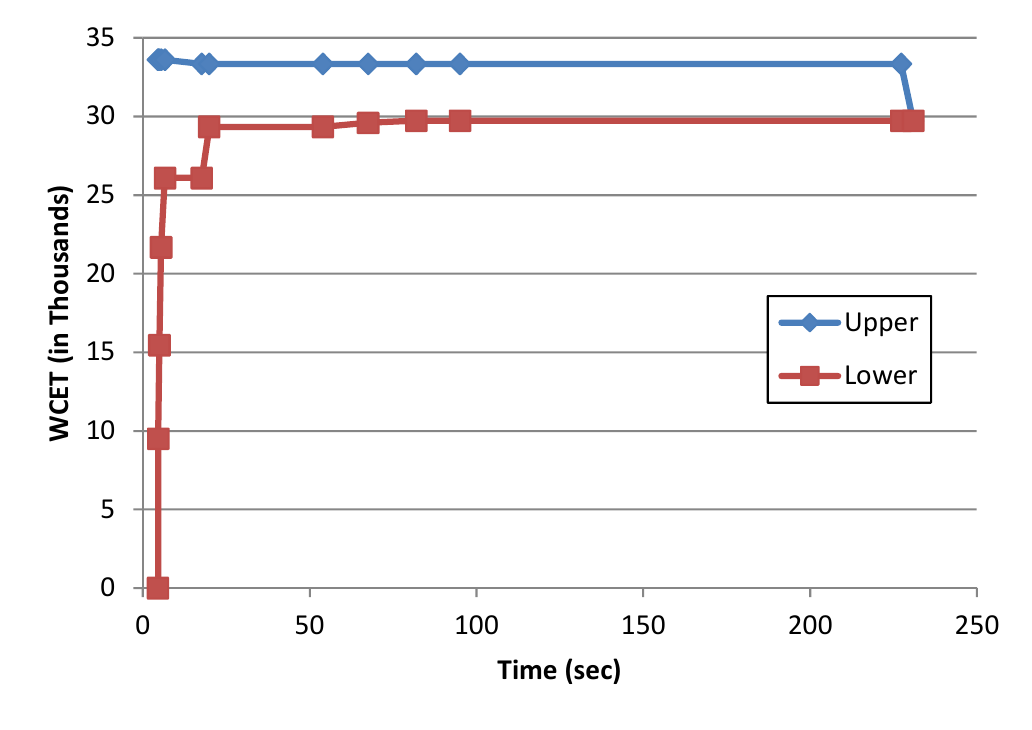}
\caption{Progressive Upper and Lower bounds over time for {\sf diskperf}}
\label{fig:trend}
\end{figure}

Finally, to observe how the upper and lower bounds incrementally
converge in our algorithm, we take a closer look the {\sf diskperf}
benchmark that best exposes this phenomenon.  Fig.~\ref{fig:trend}
shows the progressive upper and lower bound WCET of this program over
time.  The monotonicity can be clearly seen -- the lower bound always
increases and the upper bound always decreases. At any point the
algorithm is terminated, the bounds can be reported to the
user. Observe that the difference between the bounds reduces to less
than 20\% in just over 15 seconds, and when they coincide we get the
exact analysis at around 230 seconds. We noted that similar trends
were exhibited among other benchmarks as well.

\ignore{ 951

\subsection{Taint Analysis}

Table~\ref{tab:taint} shows the results of taint analysis on our benchmarks.
The column {\sf \# V} shows the total number of variables in the program, and
the columns labelled {\sf \# TV} shows the number of tainted variables. Of
course, the analysis considers variable sets, but we show only the cardinality
for presentation. 

Except for the programs {\sf nsichneu} and {\sf ssh}, the analysis produced by
AI is about 10-25\% imprecise. Again, SE fails to terminate for four out of
seven benchmarks, leaving only the imprecise result from AI to work with. In
contrary, our incremental algorithm is able to terminate with an exact analysis
for all benchmarks, and confirms the exactness in the three benchmarks that SE 
terminated on ({\sf ssh}, {\sf nsichneu}, {\sf tcas}) using fewer resources than
SE. \\

\noindent
We conclude by stating that in both Tables~\ref{tab:wcet} and~\ref{tab:taint},
our incremental algorithm comes on top in some form in every row. We surpass the exact
SE-based analyzer in performance by almost always using fewer resources, and we
surpass the AI-based analyzer in accuracy by producing more precise
analysis, together with a {\em quantification} of precision using the lower
bound. In the end, we have empirically shown that our method makes ``best use''
of the resources than both our other systems considered put together and fulfills our motive
to be an ``anytime algorithm''.
    
%

} 


\balance

\section{Related Work}
\label{sec:related}
The most related work is \cite{cerny13popl} which introduced the
original problem of quantitative analysis over a dynamic cost model.
They were the first to discuss the concept of refinement in order
to eliminate spurious analysis arising from both the infeasibility
of a path, as well as the in-optimality of the machine state.
Their main loop iterations perform abstraction refinement
in the style of the CEGAR \cite{clarke00cegar} framework.
The refinement strategy here is based on
the notion of an \emph{extremal} counterexample trace at each step, with an aim to
eliminate this trace from further consideration.  Our choice of
refinement step shares this motivation, by
choosing, in some sense, to refine the trace that maximizes the
likelihood of improvement in the analysis result.
The key technical difference between this work and ours 
is that this work refines the \emph{abstract domain}, while we refine 
the \emph{transition system}.

More specifically, 
we iteratively refine the Control Flow Graph (CFG) with appropriate splitting. 
The relationship of our refinement step to \cite{cerny13popl}'s  
is akin to that of Abstract Conflict Driven Clause Learning (ACDCL) \cite{silva13popl} 
to traditional CEGAR in the context of program verification. 
A direct tradeoff is that we need to maintain a data structure called the 
\emph{hybrid symbolic execution tree} (HSET).
But the gain is potentially significant; we quote: ``ACDCL never
changes the domain, and this immutability is crucial for efficiency
(over CEGAR), because the implementations of the abstract domain and
transformers can be highly optimized'' \cite{silva13popl}.

In the end, the algorithm of \cite{cerny13popl} is not incremental,
and does not scale to the level of our benchmarks.
The examples evaluated in \cite{cerny13popl} are very small
and can solved easily and exactly by pre-existing algorithms
such as symbolic execution.
The reason for this is partly due to the nature of CEGAR whereby
is it unclear how to ``cache'' the results of the analysis
from previous iterations, let alone in a compact form.
(In verification as opposed to analysis, we can cache the known safe states.)
Consequently, the algorithm of \cite{cerny13popl} is not progressive:
it is possible in principle to be considering the same execution path
in a nonterminating sequence of refinement.

\ignore{
\begin{itemize}
\item
Our approach possesses \emph{incremental performance}. 
The results of the present iteration are persistent, 
and can be reused in the next iteration.
In particular, paths are never analyzed twice.

\item
Our approach employs the concept of domination for effective pruning.
This is used in conjunction with a refinement strategy
which chooses to refine an abstract node that is not just un-dominated,
but one that \emph{remains} un-dominated if any other choice
of abstract node is made, thus making its refinement absolutely necessary.
\item
Our approach uses both lower and upper bound analyses, thus providing
a \emph{precision measure} for a more flexible terminating condition.
\ignore{
and effective refinement. 
We only refine an abstract node,
which not only guaranteed to be not dominated at the current iteration, but also guaranteed 
to be not  dominated in the future. 
In other words, such refinement is \emph{necessary}.
\cite{cerny13popl} choose to refine the dominating trace, given the current abstract domain.
Such trace, under different abstract domain, however, might be dominated and need not 
be considered.
}
\end{itemize}
}

The work \cite{cerny15esop} applies the concept segment-based abstraction in \cite{cerny13popl}
for high-level \wcet{} analysis. Consequently, not only have they reached a certain level 
of scalability, but also their approach can be embedded effectively
into the standard Implicit Path Enumeration Technique (\ipet) \cite{li95dac}. 
However, note importantly that in high-level \wcet{} analysis, 
as opposed to overall \wcet{} analysis, 
the timing of each basic block has been abstracted to the worst-case timing of the block, 
returned by some prior low-level analysis.  
Thus the problem no longer concerns a dynamic cost model. 
In other words, scalability is achieved partly by ignoring the issue of 
context-sensitivity raised by a dynamic cost model.

Another work related to ours is~\cite{beyer08ase} from the
BLAST~\cite{beyer07blast} line of work, which dynamically adjusts the
precision of the analysis. It carries an explicit analysis and an
abstract analysis in the form of predicates.  Then, depending on the
accumulated results, for instance when the number of explicitly tracked
values of a particular variable reaches a limit, the abstract domain is
refined by adding a predicate and the explicit analysis is abstracted by
turning it off for that variable. Our work does share a
similarity with~\cite{beyer08ase} in using both the exact and abstract
results during analysis.

The most important difference is that their work is applied on
reachability problems such as model checking and verification that are
qualitative analyses, whereas we target quantitative analyses with dynamic 
cost model. We have demonstrated clearly that for the problem domains
of interest, feasibility refinement alone is not enough.

\ignore{
Moreover, as a general problem with
predicate abstraction, their method suffers from expensive refinement
steps due to the ``globality'' of the refinement. For instance, a
variable $x$ could have reached its explicit-state limit along one path
causing the abstract domain to be refined with a predicate on $x$. But
this refinement is also considered needlessly on other paths where it is
irrelevant, i.e., where $x$ could have been not explicitly tracked.
}

Finally, we mention other related works, which share similar
motivations as our work. Many customized abstract interpreters
have been injected with some form of path-sensitivity to enhance the precision
of the analysis results. A notable example is \cite{rival07toplas}. 
There have also been work on path-sensitive algorithms (under SMT setting) 
equipped with abstract interpretation in order to  prune (a potentially
infinite number of) paths \cite{harris10popl}. However, our framework differs
significantly in the way the spines are interactively constructed. On one hand, we quickly 
refute spurious analysis from previous iteration while computing realistic lower bounds 
to exploit the new concept of domination for pruning. On the other hand, 
we can reach early termination when the spines confirm previously computed
upper bound analyses are indeed precise.

\section{Concluding Remarks}
\label{sec:conclusion}

We presented an algorithm for quantitative analysis defined
over a dynamic cost model.
The algorithm is anytime because it produces a sound analysis
after every iteration of its refinement step,
and is progressive because it eventually terminates with
an exact analysis.  Another feature is that the algorithm
computes a lower and upper bound analysis thus paving the
way for early termination, useful when the analysis
is considered good enough according to a preset level.
Finally, we show that the algorithm is incremental 
because it maintains a compact representation throughout
the refinement steps, and each new refinement step is usually
greatly assisted by the representation.
We used a well-recognized benchmark from the WCET community
to show that we can execute challenging examples.

\ignore{
We designed it so that it is incremental:
it is an ``anytime'' algorithm,
it generating an answer quickly quickly, and progressively
produces better solutions via refinement iterations.
The result of each iteration converges to an exact analysis when given
an unlimited resource budget.  
We finally give evidence that a new level of practicality is achieved
by an evaluation on a realistic collection of benchmarks.
}

\ignore{
We presented an algorithm for quantitative analysis that produces
results of increasing precision in incremental steps. Providing an
alternative to the CEGAR-based methods currently used, our method uses
symbolic execution to refine the hybrid graph structure instead of the
abstract domain, with several desirable features as a result.

A first feature is that a sound analysis
is obtained after any number of iterations, 
and exact precision is obtained eventually.
A second feature is that our analysis comprises of both lower and upper
bounds, which allow user-definable levels of acceptable precision and
the possibility of early termination once the level is reached.
A third feature is that the algorithm is equipped with a concept of
domination which can prune the search space.
A fourth feature is that the algorithm is both incremental and goal-directed in
its refinement process, and therefore pruning is, arguably, often effective.

Finally, our realistic benchmarks, on two complementary kinds (backward and forward)
of analyses, show that our algorithm outperforms in almost all respects.
On examples for which a non-iterative exact analyzer can terminate within a budget,
our algorithm almost always utilizes less memory and time.
For examples on which no known exact analyzer can terminate within that budget,
our algorithm not only produces a more accurate analysis than an abstract analysis,
but it quantifies the precision using upper and lower bounds.

In summary, we believe that fundamental alternatives to CEGAR-based
methods provide several advantages in the context of quantitative
analysis. This paper presented once such method and showed empirically
that it is both viable and competitive with current methods.

}

%




\bibliographystyle{plain}
\bibliography{references}

\begin{thebibliography}{26}
\providecommand{\natexlab}[1]{#1}
\providecommand{\url}[1]{\texttt{#1}}
\expandafter\ifx\csname urlstyle\endcsname\relax
  \providecommand{\doi}[1]{doi: #1}\else
  \providecommand{\doi}{doi: \begingroup \urlstyle{rm}\Url}\fi

\bibitem[Banerjee et~al.(2013)Banerjee, Chattopadhyay, and
  Roychoudhury]{banerjee13rtss}
A.~Banerjee, S.~Chattopadhyay, and A.~Roychoudhury.
\newblock Static analysis driven cache performance testing.
\newblock In \emph{RTSS}, pages 319--329, 2013.

\bibitem[Beyer(2014)]{SVCOMP14}
D.~Beyer.
\newblock Third competition on software verification.
\newblock In \emph{TACAS}, 2014.

\bibitem[Beyer et~al.(2007)Beyer, Henzinger, Jhala, and Majumdar]{beyer07blast}
D.~Beyer, T.~Henzinger, R.~Jhala, and R.~Majumdar.
\newblock The {S}oftware {M}odel {C}hecker {BLAST}.
\newblock \emph{Int. J. STTT}, 9:\penalty0 505--525, 2007.

\bibitem[Beyer et~al.(2008)Beyer, Henzinger, and Theoduloz]{beyer08ase}
D.~Beyer, T.~A. Henzinger, and G.~Theoduloz.
\newblock Program analysis with dynamic precision adjustment.
\newblock In \emph{ASE}, 2008.

\bibitem[Boddy(1991)]{boddy91aaai}
M.~Boddy.
\newblock Anytime problem solving using dynamic programming.
\newblock In \emph{AAAI}, pages 738--743, 1991.

\bibitem[Cerny et~al.(2013)Cerny, Henzinger, and Radhakrishna]{cerny13popl}
P.~Cerny, T.~A. Henzinger, and A.~Radhakrishna.
\newblock Quantitative abstraction refinement.
\newblock In \emph{POPL}, pages 115--128, 2013.

\bibitem[Cerny et~al.(2015)Cerny, Henzinger, Kovacs, Radhakrishna, and
  Zwirchmayr]{cerny15esop}
P.~Cerny, T.~A. Henzinger, L.~Kovacs, A.~Radhakrishna, and J.~Zwirchmayr.
\newblock Segment abstraction for worst-case execution time analysis.
\newblock In \emph{ESOP}, pages 105--131, 2015.

\bibitem[Chu and Jaffar(2011)]{chu11emsoft}
D.~H. Chu and J.~Jaffar.
\newblock Symbolic simulation on complicated loops for wcet path analysis.
\newblock In \emph{EMSOFT}, 2011.

\bibitem[Cimatti et~al.(2008)Cimatti, Griggio, and Sebastiani]{Cimatti-Tacas08}
A.~Cimatti, A.~Griggio, and R.~Sebastiani.
\newblock Efficient interpolant generation in satisfiability modulo theories.
\newblock In \emph{TACAS'08}, pages 397--412, 2008.

\bibitem[Clarke et~al.(2000)Clarke, Grumberg, Jha, Lu, and
  Veith]{clarke00cegar}
E.~Clarke, O.~Grumberg, S.~Jha, Y.~Lu, and H.~Veith.
\newblock {C}ounter{E}xample-{G}uided {A}bstraction {R}efinement.
\newblock In \emph{CAV}, 2000.

\bibitem[Craig(1955)]{craig55interpolant}
W.~Craig.
\newblock Three uses of {Herbrand}-{Gentzen} theorem in relating model theory
  and proof theory.
\newblock \emph{Journal of Symbolic Computation}, 22, 1955.

\bibitem[Dijkstra(1975)]{dijkstrawp}
E.~W. Dijkstra.
\newblock Guarded commands, nondeterminacy and formal derivation of programs.
\newblock \emph{Commun. ACM}, 1975.

\bibitem[D'Silva et~al.(2013)D'Silva, Haller, and Kroening]{silva13popl}
V.~D'Silva, L.~Haller, and D.~Kroening.
\newblock Abstract conflict driven learning.
\newblock In \emph{POPL}, pages 143--154, 2013.

\bibitem[Harris et~al.(2010)Harris, Sankaranarayanan, Ivan\v{c}i\'{c}, and
  Gupta]{harris10popl}
W.~R. Harris, S.~Sankaranarayanan, F.~Ivan\v{c}i\'{c}, and A.~Gupta.
\newblock Program analysis via satisfiability modulo path programs.
\newblock In \emph{POPL}, pages 71--82, 2010.

\bibitem[Jaffar et~al.(2008)Jaffar, Santosa, and Voicu]{jaffar08aaai}
J.~Jaffar, A.~E. Santosa, and R.~Voicu.
\newblock Efficient memoization for dynamic programming with ad-hoc
  constraints.
\newblock In \emph{AAAI}, 2008.

\bibitem[Jaffar et~al.(2009)Jaffar, Santosa, and Voicu]{jaffar09intp}
J.~Jaffar, A.~E. Santosa, and R.~Voicu.
\newblock An interpolation method for {CLP} traversal.
\newblock In \emph{15th CP, LNCS 5732}, 2009.

\bibitem[{J}affar et~al.(2012){J}affar, {M}urali, {N}avas, and
  {S}antosa]{jaffar12sas}
J.~{J}affar, V.~{M}urali, J.~{N}avas, and A.~{S}antosa.
\newblock Path sensitive backward analysis.
\newblock In \emph{SAS}, 2012.

\bibitem[Li and Malik(1995)]{li95dac}
Y.-T.~S. Li and S.~Malik.
\newblock Performance analysis of embedded software using implicit path
  enumeration.
\newblock In \emph{DAC}, 1995.

\bibitem[M\"{a}lardalen()]{malardalenbenchmark}
M\"{a}lardalen.
\newblock M\"{a}lardalen {WCET} research group benchmarks.
\newblock URL {\tt
  h\-t\-t\-p://www.m\-r\-t\-c.m\-d\-h.se/pro\-jects/w\-c\-e\-t/bench\-marks.html},
  2006.

\bibitem[McMillan(2010)]{mcmillan10cav}
K.~L. McMillan.
\newblock Lazy annotation for program testing and verification.
\newblock In \emph{CAV}, 2010.

\bibitem[Prantl et~al.(2008)Prantl, Schordan, and Knoop]{TuBound}
A.~Prantl, M.~Schordan, and J.~Knoop.
\newblock Tu{B}ound -- a conceptually new tool for worst-case execution time
  analysis.
\newblock In \emph{WCET}, 2008.

\bibitem[Puschner and Burns(2000)]{WCETOverview00}
P.~Puschner and A.~Burns.
\newblock A review of worst-case execution-time analysis.
\newblock \emph{Journal of Real-Time Systems}, 2000.

\bibitem[Rival and Mauborgne(2007)]{rival07toplas}
X.~Rival and L.~Mauborgne.
\newblock The trace partitioning abstract domain.
\newblock \emph{ACM Trans. Program. Lang. Syst.}, 29\penalty0 (5), Aug. 2007.

\bibitem[Theiling et~al.(2000)Theiling, Ferdinand, and Wilhelm]{theiling00wcet}
H.~Theiling, C.~Ferdinand, and R.~Wilhelm.
\newblock Fast and precise {WCET} prediction by seperate cache and path
  analyses.
\newblock \emph{Real-Time Systems}, 18\penalty0 (2/3):\penalty0 157--179, May
  2000.

\bibitem[Tiwari et~al.(1994)Tiwari, Malik, and Wolfe]{tiwari94iccad}
V.~Tiwari, S.~Malik, and A.~Wolfe.
\newblock Power analysis of embedded software: A first step towards software
  power minimization.
\newblock In \emph{ICCAD}, pages 384--390, 1994.

\bibitem[Wilhelm et~al.(2008)Wilhelm, Engblom, Ermedahl, Holsti, Thesing,
  Whalley, Bernat, Ferdinand, Heckmann, Mitra, Mueller, Puaut, Puschner,
  Staschulat, and Stenstr\"{o}m]{WCETOverview08}
R.~Wilhelm, J.~Engblom, A.~Ermedahl, N.~Holsti, S.~Thesing, D.~Whalley,
  G.~Bernat, C.~Ferdinand, R.~Heckmann, T.~Mitra, F.~Mueller, I.~Puaut,
  P.~Puschner, J.~Staschulat, and P.~Stenstr\"{o}m.
\newblock The worst-case execution-time problem---overview of methods and
  survey of tools.
\newblock \emph{Trans. on Embedded Computing Sys.}, 2008.

\end{thebibliography}

\end{document}